\documentclass[aps,prd,preprint,nofootinbib,longbibliography,superscriptaddress,eqsecnum,amsmath,amsfonts,amssymb]{revtex4-2}
\usepackage{graphicx} 
\usepackage{physics}
\usepackage{tikz}
\usepackage{hyperref}
\usepackage{bbm}

\newcommand\nn{\nonumber}

\newcommand{\be}{\begin{equation}}
\newcommand{\ee}{\end{equation}}
\newcommand{\bea}{\setlength\arraycolsep{2pt} \begin{eqnarray}}
\newcommand{\eea}{\end{eqnarray}}

\def\fft#1#2{{\frac{#1}{#2}}}

\def\0{{\sst{(0)}}}
\def\1{{\sst{(1)}}}
\def\2{{\sst{(2)}}}
\def\3{{\sst{(3)}}}
\def\4{{\sst{(4)}}}
\def\5{{\sst{(5)}}}
\def\6{{\sst{(6)}}}
\def\7{{\sst{(7)}}}
\def\8{{\sst{(8)}}}
\def\sst#1{{\scriptscriptstyle #1}}

\allowdisplaybreaks[1]

\begin{document}
\title{Consistent Four-derivative Heterotic Truncations and the Kerr-Sen Solution}
\preprint{USTC-ICTS/PCFT-25-34}
\date{\today}

\author{Minghao Xia}\email{1024210011@tju.edu.cn}
\affiliation{Center for Joint Quantum Studies and Department of Physics, School of Science, Tianjin University, Tianjin 300350, China}

\author{Liang Ma}\email{maliang0@tju.edu.cn}
\affiliation{Center for Joint Quantum Studies and Department of Physics, School of Science, Tianjin University, Tianjin 300350, China}

\author{Yi Pang}\email{pangyi1@tju.edu.cn}
\affiliation{Center for Joint Quantum Studies and Department of Physics, School of Science, Tianjin University, Tianjin 300350, China}
\affiliation{Peng Huanwu Center for Fundamental Theory, Hefei, Anhui 230026, China}

\author{Robert J. Saskowski}\email{robert\_saskowski@tju.edu.cn}
\affiliation{Center for Joint Quantum Studies and Department of Physics, School of Science, Tianjin University, Tianjin 300350, China}

\begin{abstract}
    Four-derivative heterotic supergravity (without gauge fields) reduced on a $p$-dimensional torus leads to half-maximal supergravity coupled to $p$ vector multiplets, and it is known that removing the vector multiplets is a consistent truncation of the theory. We find a new consistent truncation of four-derivative heterotic supergravity on a torus that keeps the vector multiplets and precisely reproduces the bosonic action of heterotic supergravity (with heterotic gauge fields). We show that both truncations have an $O(d+p,d)$ symmetry when reduced on a $d$-dimensional torus and demonstrate how this embeds in the $O(d+p,d+p)$ symmetry that one gets from reducing on a $(d+p)$-dimensional torus without truncation. We then use our new truncation to obtain four-derivative corrections to the Kerr-Sen solution and compute thermodynamic quantities and multipole moments. Finally, we compare the Kerr-Sen solutions of the actions corresponding to the two different choices of truncation with the Kerr solution, the Kerr-Newman solution, and each other, and show that they have distinct four-derivative multipole structures.
\end{abstract}

\maketitle
\newpage

\tableofcontents

\section{Introduction and Summary}
String theory is exceptional in that the fundamental objects of the theory have an extended spatial dimension. In particular, if we put the theory on a torus, strings can both move around the compact dimensions or wrap them, which leads to a symmetry that mixes the momentum and winding modes, \emph{i.e.}, T-duality. Thus, reducing string theory on a $d$-dimensional torus leads to a ``hidden'' $O(d,d;\mathbb Z)$ symmetry appearing in the reduced theory. For the heterotic string, there are also gauge fields living in either $E_8\times E_8$ or $SO(32)$. Dimensional reduction on a torus will preserve at most the maximal abelian subgroup of the gauge group, namely $U(1)^{16}$, which enhances the $O(d,d;\mathbb Z)$ symmetry to $O(d+16,d;\mathbb Z)$. More generally, if only a $U(1)^p$ subgroup is preserved, then the symmetry is enhanced to $O(d+p,d;\mathbb Z)$. In the low-energy limit, string theory gives way to supergravity, and, as we forget about the quantization of charges in the semiclassical limit, the $O(d+p,d;\mathbb Z)$ symmetry is further enhanced to $O(d+p,d;\mathbb R)$ symmetry, wherein the scalars parametrize a $O(d+p,d)/(O(d+p-1,1)\times O(d-1,1))$ coset,\footnote{Note that this is in Lorentzian signature since we will be interested in compactifying time.} the gauge fields transform as a vector under $O(d+p,d)$, and the fermions transform under $O(d+p-1,1)\times O(d-1,1)$, while the metric, three-form NS flux, and dilaton all transform as scalars.

Such a symmetry in supergravity can be used to great effect to generate new solutions~\mbox{\cite{Veneziano:1991ek,Meissner:1991zj,Sen:1991zi,Sen:1991cn,Gasperini:1991qy,Hassan:1991mq,Sen:1992ua,Cvetic:1995sz,Cvetic:1995kv,Cvetic:1996xz}}. In particular, having an abelian isometry is (mathematically) equivalent to compactification on a torus, so any heterotic supergravity solution with $p$ abelian gauge fields and a $U(1)^d$ isometry will naturally have an $O(d+p,d)$ symmetry that can be used to generate new solutions. In particular, time-independent solutions have such an isometry; in Euclidean signature, this is genuinely a $U(1)$ isometry with periodicity set by the inverse temperature, but it is also perfectly (mathematically) consistent to Wick rotate to Lorentzian signature and treat the $\mathbb R$ time isometry as an (infinite radius) $S^1$ reduction. 

The $O(d+p,d)$ symmetry is expected to persist to all orders in the $\alpha'$ expansion~\cite{Sen:1991zi,Hohm:2014sxa}, which means that the solution-generating procedure can also be applied to generate higher-derivative corrections~\cite{Hu:2025aji}. Much work on four-derivative heterotic supergravity has truncated the gauge fields, and so it is natural to ask if it is possible to reinstate these gauge fields without redoing all these calculations from scratch. As we will see, we can indeed reinstate the heterotic gauge fields. The presence of an ${O(d+p,d)}$ symmetry places considerable constraint on the form of the higher-derivative terms, as evidenced both by direct construction~\cite{Godazgar:2013bja,Hohm:2015doa,Marques:2015vua,Baron:2017dvb,Garousi:2019wgz,Garousi:2019mca,Garousi:2020gio,Codina:2020kvj,David:2021jqn,David:2022jcl,Ozkan:2024euj} and double field theory~\cite{Marques:2015vua,Hohm:2013jaa,Bedoya:2014pma,Hohm:2014xsa,Coimbra:2014qaa,Lee:2015kba,Baron:2017dvb,Lescano:2021guc}.\footnote{However, note that such DFT formulations break down at higher orders in $\alpha'$, namely at $\mathcal O(\alpha'^3)$~\cite{Hronek:2020xxi,Hsia:2024kpi}.}\footnote{See also~\cite{Lunin:2024vsx} for an example of $\alpha'$-corrections computed directly within the double field theory framework.} In particular, heterotic supergravity without gauge fields has only an $O(d,d)$ symmetry, and this fixes the four-derivative action uniquely up to field redefinitions and a choice of worldsheet parity~\cite{Baron:2017dvb,Marques:2015vua}. When the gauge fields are included, there are actually two distinct $O(d+p,d)$ invariant actions at the four-derivative level~\cite{Baron:2017dvb}, one that corresponds precisely to heterotic supergravity with gauge fields~\cite{Bergshoeff:1988nn,Bergshoeff:1989de} and one that corresponds to the truncation found in~\cite{Liu:2023fqq}.

Within this context, time-independent solutions such as the Kerr solution can be rotated using $O(2,1)$ to turn on a $U(1)$ charge. In four dimensions, this procedure generates the Kerr-Sen solution~\cite{Sen:1991zi}, which is the unique stationary axisymmetric solution to four-dimensional heterotic supergravity with one abelian gauge field~\cite{Rogatko:2010hf}. Recently, Ref.~\cite{Hu:2025aji} performed the $O(2,1)$ boost procedure to find the four-derivative corrections to the Kerr-Sen solution for the theory corresponding to the truncation in~\cite{Liu:2023fqq}. Of note, the main complication that arises at the four-derivative level is that the reduced action is not automatically $O(d+p,d)$ invariant, unlike the two-derivative case. Instead, one must perform field redefinitions to make the $O(d+p,d)$ symmetry manifest~\cite{Eloy:2020dko,Elgood:2020xwu,Ortin:2020xdm,Jayaprakash:2024xlr}. Moreover, since the authors used the results for $O(d,d)$, they embedded $O(2,1)$ into $O(2,2)$ as a trick to recover the gauge field. In this paper, we will be interested in extending these results to the other $O(d+p,d)$ invariant action, \emph{i.e.}, heterotic supergravity with gauge fields.

Of course, as physicists, we must always keep in mind the potential experimental tests of our results. In the present context, since we are working in four dimensions, the NS two-form can be Hodge dualized to an axion that combines with the dilaton to form a complex scalar field, which may potentially be viewed as a sort of complex inflaton~\mbox{\cite{Sonner:2006yn,Catena:2007jf,Linde:1991km}}. Likewise, one may consider the $U(1)$ field to be either the photon of the Standard Model or of the hidden sector. If we consider it to be the photon, then we should compare with the Kerr-Newman solution of Einstein-Maxwell theory, and if we consider it to be the dark photon, then we should compare with the Kerr solution. Either way, as was the case in~\cite{Hu:2025aji}, the Kerr-Sen solution has a distinct astrophysical signature. The Kerr-Sen solution is a natural candidate to compare with experimental observations and much work has already been devoted to this endeavor~\cite{Campbell:1992hc,Dastan:2016bfy,Cvetic:2017zde,Guo:2019lur,Narang:2020bgo,Xavier:2020egv,Shahzadi:2022rzq,DellaMonica:2023ydm,Sahoo:2023czj,Feng:2024iqj}. However, with the advent of precision gravitational wave measurement and, in particular, with upcoming extremal mass ratio inspiral experiments~\cite{LISA:2017pwj,Barack:2006pq,Cardoso:2016ryw,Babak:2017tow,Ryan:1995wh,Krishnendu:2018nqa}, a good candidate for comparison is the multipole moments. At the two-derivative level, the gravitational multipole moments are the same for Kerr, Kerr-Newman, and Kerr-Sen black holes, but, as was shown in~\cite{Hu:2025aji}, they differ at the four-derivative level, which motivates us to compute the four-derivative corrections to the multipole moments, which we will find are distinct.

\subsection{Summary}
$(d+p)$-dimensional heterotic supergravity (without gauge fields) reduced on $T^p$ leads to $d$-dimensional half-maximal supergravity coupled to $p$ vector multiplets. Schematically, this decomposition takes the form
\begin{equation}
    (\hat g_{MN},B_{MN},\phi)\to(g_{\mu\nu},b_{\mu\nu},A^{(+)\,i}_\mu,\varphi)+(A^{(-)\,i}_\mu,g_{ij}, b_{ij}),
\end{equation}
for the bosonic fields. At the four-derivative level, one possible consistent truncation is to remove the vector multiplets by truncating $A^{(-)\,i}$, $g_{ij}$, and $b_{ij}$, which yields the action~\cite{Liu:2023fqq}
\begin{align}
    e^{-1}\mathcal L^{(+)}_d&=e^{-2\varphi}\left\{R+4(\partial\varphi)^2-\frac{1}{12}h^2-\frac{1}{4}\qty(\mathcal F^i)^2\nn\right.\\
    &\qquad\qquad\left.+\frac{\alpha'}{8}\qty[R(\omega_-)^2_{\mu\nu\rho\sigma}-R(\omega_-)^{\mu\nu\rho\sigma}\mathcal F^i_{\mu\nu}\mathcal F^i_{\rho\sigma}-\frac{1}{2}\mathcal F^i\mathcal F^i\mathcal F^j\mathcal F^j+\frac{1}{2}\mathcal F^i\mathcal F^j\mathcal F^i\mathcal F^j]\right\},\label{eq:TruncP}
\end{align}
and preserves half-maximal supersymmetry. We find a new truncation that corresponds to removing $A^{(+)\,i}$, $g_{ij}$, and $b_{ij}$, which yields the action
\begin{align}
    e^{-1}\mathcal L^{(-)}_d&=e^{-2\varphi}\left\{R+4(\partial\varphi)^2-\frac{1}{12}h^2-\frac{1}{4}\qty(\mathcal F^i)^2\right.\nn\\
    &\qquad\qquad\left.+\frac{\alpha'}{8}\qty[R(\omega_-)^2_{\mu\nu\rho\sigma}-\frac{1}{2}\mathcal F^i\mathcal F^i\mathcal F^j\mathcal F^j+\mathcal F^i\mathcal F^j\mathcal F^i\mathcal F^j-\frac{1}{2}\qty(\mathcal F^2_{ij})^2]\right\},\label{eq:TruncM}
\end{align}
and verify the consistency of this truncation. Moreover, the action~\eqref{eq:TruncM} matches the action of Bergshoeff and de Roo for heterotic supergravity with gauge fields~\cite{Bergshoeff:1988nn,Bergshoeff:1989de}. Clearly, this truncation does not preserve half-maximal supersymmetry, but we show that it preserves $\mathcal N=1$ supersymmetry in four dimensions.

If we start in $D+p$ dimensions, reduce on $T^{p}$, and apply one of these two truncations along $T^p$, then we are left with $D$-dimensional heterotic supergravity coupled to $p$ vector fields. We can then further reduce this vector-coupled supergravity on $T^d$, without further truncation. We show that either choice of truncation leads to an $O(d+p,d)$ symmetry that is naturally embedded inside the $O(d+p,d+p)$ symmetry that appears in the untruncated theory reduced on $T^{d+p}$. We also explicitly construct the isomorphism between the two copies of $O(d+p,d)$, corresponding to the two choices of truncation.

We then make use of this embedding to find the four-derivative corrections to the Kerr-Sen solution using the boost procedure of~\cite{Hu:2025aji}. This involves embedding the four-dimensional four-derivative corrected Kerr solution into five dimensions, reducing to three dimensions, redefining to the $O(2,1)$ invariant field redefinition frame, performing the Kerr-Sen $O(2,1)$ boost, field redefining back and uplifting to five dimensions, and, finally, reducing back to four dimensions. The use of the five-dimensional uplift as an intermediary step is equivalent to working with heterotic supergravity without truncating the gauge fields. The corrections for the truncation~\eqref{eq:TruncP} were already computed in~\cite{Hu:2025aji}, but the result for the truncation~\eqref{eq:TruncM} is entirely novel. 

We then compute the thermodynamics and multipole moments of the corrected Kerr-Sen solutions. The Kerr-Sen multipole moments match those of the Kerr(-Newman) black hole at the two-derivative level, so we must compare at the four-derivative level. In particular, we find that both Kerr-Sen solutions have distinct four-derivative gravitational multipole structures from the Kerr solution. That is to say, if the $U(1)$ gauge field were viewed as a hidden sector field, then we could experimentally distinguish either Kerr-Sen solution from the Kerr solution. On the other hand, if we consider the $U(1)$ gauge field to be the photon of the Standard Model, then we should compare with the Kerr-Newman solution. We find that both Kerr-Sen solutions have distinct four-derivative gravitational and electromagnetic multipole structures from Kerr-Newman black holes, even with the most general choice of four-derivative corrections to the Einstein-Maxwell theory.

The rest of this paper is organized as follows. In Section~\ref{sec:genDims}, we review the torus reduction and find a new consistent four-derivative truncation. We then show how $O(d+p,d)$ embeds inside of $O(d+p,d+p)$. In Section~\ref{sec:4D}, we perform the $O(2,1)$ boost procedure for both truncations to obtain four-derivative corrected Kerr-Sen solutions. We then compute the thermodynamics and multipole moments and compare with the Kerr and Kerr-Newman solutions. Finally, we conclude in Section~\ref{sec:disc} and discuss future directions. Appendix~\ref{app:susy} analyzes the supersymmetry of the two truncations in four dimensions.

\section{General Dimensions}\label{sec:genDims}
Heterotic supergravity is a half-maximal supergravity theory consisting of an $\mathcal N=(1,0)$ gravity multiplet $(\hat g_{MN},\psi_M,B_{MN},\lambda,\phi)$ coupled to vector multiplets $(\mathcal A_M^{\mathbbm{i}},\chi^\mathbbm{i})$, transforming under either $E_8\times E_8$ or $SO(32)$. The field content is thus the metric $\hat g$, the gravitino $\psi$, the NS two-form $B$, the dilatino $\lambda$, the dilaton $\phi$, the gauge fields $\mathcal A^\mathbbm{i}$, and the gaugini $\chi^\mathbbm{i}$. Formally, this theory is defined in ten dimensions, but, at least at the bosonic level, we can just as well choose to put the theory in $D$ dimensions. If $D<10$, this is morally equivalent to doing a trivial reduction on $T^{10-D}$. The four-derivative action in Bergshoeff-de Roo (BdR) form is given by~\cite{Bergshoeff:1988nn,Bergshoeff:1989de}\footnote{Note that in comparing with~\cite{Bergshoeff:1988nn,Bergshoeff:1989de}, one must flip $\Omega\to-\Omega$, $B\to-B$, and $\alpha'\to-\alpha'$.}
\begin{align}
    e^{-1}\mathcal{L}_D&=e^{-2\phi}\left\{R(\Omega)+4(\partial\phi)^2-\frac{1}{12}\tilde H^2-\frac{1}{4}\qty(\mathcal F^\mathbbm{i})^2\right.\nn\\
    &\quad\left.+\frac{\alpha'}{8}\qty[R_{MNPQ}(\Omega_-)^2-\frac{1}{2}\mathcal F_{MN}^\mathbbm{i}\mathcal F_{NP}^\mathbbm{i}\mathcal F_{PQ}^\mathbbm{j}\mathcal F_{QM}^\mathbbm{j}+\mathcal F_{MN}^\mathbbm{i}\mathcal F_{NP}^\mathbbm{j}\mathcal F_{PQ}^\mathbbm{i}\mathcal F_{QM}^\mathbbm{j}-\frac{1}{2}\qty(\mathcal F^\mathbbm{i}_{MN}\mathcal F^{\mathbbm j\,MN})^2]\right\},\label{eq:BdR4d}
\end{align}
where $R$ is the Ricci scalar and $\Omega$ the Levi-Civita spin connection, and we have defined
\begin{equation}
    \tilde H= \dd B+\frac{1}{2}\mathcal A^\mathbbm{i}\land \mathcal F^\mathbbm{i}+\frac{\alpha'}{4}\omega_{3L}(\Omega_-),
\label{eq:Htilde}
\end{equation}
to be the three-form NS flux. Here, $\mathcal F^\mathbbm{i}$ is the field strength for $\mathcal A^\mathbbm{i}$, and we have introduced the torsionful spin connection
\begin{equation}
    \Omega_\pm=\Omega\pm\fft12 \tilde H.
\end{equation}
The corresponding curvature is given by
\begin{equation}
    R(\Omega_\pm)=\dd\Omega_\pm+\Omega_\pm\wedge\Omega_\pm,
\end{equation}
while the Lorentz-Chern-Simons form is defined by
\begin{equation}
    \omega_{3L}(\Omega_\pm)=\Tr\left(\Omega_\pm\wedge \dd\Omega_\pm+\fft23\Omega_\pm\wedge\Omega_\pm\wedge\Omega_\pm\right).
\end{equation}
Note that the three-form flux satisfies the Bianchi identity
\begin{equation}
    \dd\tilde H=\frac{1}{2}F^\mathbbm{i}\land F^\mathbbm{i} + \frac{\alpha'}{4}\Tr \qty[R(\Omega_-)\land R(\Omega_-)].\label{eq:Bianchi}
\end{equation}

Truncating away the heterotic gauge fields leaves us with the simpler bosonic action
\begin{equation}
    e^{-1}\mathcal{L}_D=e^{-2\phi}\qty[R(\Omega)+4(\partial\phi)^2-\frac{1}{12}\tilde H^2+\frac{\alpha'}{8}R_{MNPQ}(\Omega_-)^2],\qquad \tilde H= \dd B+\frac{\alpha'}{4}\omega_{3L}(\Omega_-),\label{eq:noGaugeFields}
\end{equation}
which we will be interested in reducing on a torus.

\subsection{The Torus Reduction}
Following~\cite{Jayaprakash:2024xlr}, we may start in $D=d+p$ dimensions and reduce \eqref{eq:noGaugeFields} on a torus $T^p$ using the standard ansatz
\begin{align}
    \dd\hat s^2&=g_{\mu\nu}\dd x^\mu\dd x^\nu+g_{ij}\eta^i\eta^j,\qquad \eta^i=\dd y^i+A^i_\mu\dd x^\mu,\nn\\
    B&=\frac{1}{2}b_{\mu\nu}\dd x^\mu\land\dd x^\nu+B_{\mu i}\dd x^\mu\land\eta^i+\frac{1}{2}b_{ij}\eta^i\land\eta^j,\nn\\
    \phi&=\varphi+\frac{1}{4}\log|\det g_{ij}|.\label{eq:redAnsatz}
\end{align}
$A_\mu^i$ is a principal $U(1)^p$ connection with curvature $F^i=\dd A^i$, $g_{\mu\nu}$ the ambient $d$-dimensional metric, and $g_{ij}$ a symmetric matrix of scalars. Note that $x^\mu$ are coordinates in the $d$-dimensional base space, while $y^i$ are coordinates on $T^p$. It then follows that
\begin{equation}
    H=\dd B=\frac{1}{6}h_{\mu\nu\rho}\dd x^\mu\land\dd x^\nu\land\dd x^\rho+\frac{1}{2}\tilde G_{\mu\nu i}\dd x^\mu\land\dd x^\nu\land\eta^i+\frac{1}{2}\dd b_{ij}\eta^i\land\eta^j,
\end{equation}
where
\begin{align}
    h=\dd b-B_i\land F^i,\qquad \tilde G_i=G_i-b_{ij}F^j,\qquad G_i=\dd B_i\,.
\end{align}
Notably, $B_{i}$ does not transform merely as a gauge field but rather has extra structure arising from the two-group gauge symmetry of the $B$ field. The scalar kinetic terms are given by
\begin{align}
    P^{(-+)}_{ij}=P^{(+-)}_{ji}=\frac{1}{2}\dd(g_{ij}+b_{ij}).
\end{align}
Choosing a vielbein\footnote{Looking ahead, we will be interested in reducing along the time direction, hence why the scalar ``vielbein'' is the Lorentzian one in our conventions.} $g_{\mu\nu}=e_\mu^\alpha e_\nu^\beta\delta_{\alpha\beta}$ and a scalar ``vielbein'' $g_{ij}=e_i^a e_j^b\eta_{ab}$, there are also composite $O(p-1,1)_-\times O(1,p-1)_+$ connections
\begin{align}
    Q^{(\pm\pm)}_{ij}&=\frac{1}{2}\qty(e^T\dd e-\dd e^T\,e\mp\dd b)_{ij}.
\end{align}

This yields an effective $d$-dimensional action describing a half-maximal gravity multiplet coupled to $p$ vector multiplets. From the supersymmetry variations, one sees that $F^{(+)\,i}$ lies in the half-maximal graviton multiplet and $(F^{(-)\,i},P^{(-+)}_{ij})$ constitute the vector multiplets~\cite{Liu:2023fqq}, where
\begin{equation}
    F^{(\pm)i}=F^i\pm g^{ij}\tilde G_j.
\end{equation}

The resulting Lagrangian is not manifestly $O(d,d)$ invariant and has derivatives of field strengths.\footnote{As a slight abuse of language, we refer to $\partial_\mu\varphi$, $\partial_\mu g_{ij}$, and $\partial_\mu b_{ij}$ as field strengths.} In particular, we redefine~\cite{Jayaprakash:2024xlr,Cai:2025yyv}\footnote{Note that our conventions match those of~\cite{Hu:2025aji}, which differ from those of~\cite{Liu:2023fqq,Jayaprakash:2024xlr,Cai:2025yyv} by a flip of worldsheet parity, $B\to-B$, which is equivalent to $b\to-b$, $B_i\to-B_i$, $b_{ij}\to-b_{ij}$.}
\begin{align}
    \delta g_{ij}&=\frac{1}{16}F^{(+)}_{\mu\nu\,i}F^{(+)}_{\mu\nu\,j}+\frac{1}{2}P_\gamma{}^{a}{}_i P_\gamma{}^{a}{}_j,\nn\\
    \delta B_{\mu i}&=\frac{1}{4}\qty(\frac{1}{2}\omega_{-\mu}^{\alpha\beta}F^{(+)}_{\alpha\beta i}+F^{(-)\,j}_{\mu\nu}P^{(-+)\,\nu}{}_{ji}),
\end{align}
in order to make the $O(d,d)$ invariance manifest, and then redefine
\begin{align}
    \delta g_{\mu\nu}&=F_{\mu\lambda}^{(+)\,a}F_{\nu\lambda}^{(+)\,a}+F_{\mu\lambda}^{(-)\,a}F_{\nu\lambda}^{(-)\,a}+4P_\mu^{(-+)\,ab}P_\nu^{(-+)\,ab},\nn\\
    \delta\varphi&=\fft14F_{\mu\nu}^{(+)\,a}F_{\mu\nu}^{(+)\,a}+\fft14F_{\mu\nu}^{(-)\,a}F_{\mu\nu}^{(-)\,a}+P_\mu^{(-+)\,ab}P_\mu^{(-+)\,ab},\nn\\
    \delta A_\mu^i&=\dfrac{1}{4}e^i_a\qty(4F_{\mu\nu}^{(-)\,b}P_\nu^{(-+)\,ba}-h_{\mu}{}^{\nu\lambda}F_{\nu\lambda}^{(+)\,a}),\nn\\
    \delta B_{\mu i}&=\dfrac{1}{4}\qty(e_{ia}-b_{ij}e^j_a)\qty(4F_{\mu\nu}^{(-)\,b}P_\nu^{(-+)\,ba}-h_{\mu}{}^{\nu\lambda}F_{\nu\lambda}^{(+)\,a}),\nn\\
    \delta b_{\mu\nu}&=B_{[\nu|i}\delta A_{|\mu]}^i,\label{eq:secondFieldsRedefs}
\end{align}
to remove explicit derivatives of field strengths. The resulting Lagrangian is given by~\cite{Jayaprakash:2024xlr}
\begin{equation}
    \mathcal L_d=\mathcal L_{2\partial}+\frac{\alpha'}{8}\qty(\mathcal L_1+\mathcal L_2+\mathcal L_3),\label{eq:redAction}
\end{equation}
where the two-derivative action is given by
\begin{align}
    e^{-1}\mathcal L_{2\partial}&=e^{-2\varphi}\left[R(\omega)+4(\partial\varphi)^2-\frac{1}{12}h^2-\frac{1}{4}\qty(g_{ij}F^i_{\mu\nu}F^{\mu\nu j}+g^{ij}\tilde G_{\mu\nu i}\tilde G^{\mu\nu}_ j)\right.\nn\\
    &\qquad\qquad\left.-\frac{1}{4}\Tr\qty(g^{-1}\partial_\mu g g^{-1}\partial^\mu g-g^{-1}\partial_\mu b g^{-1}\partial^\mu b)\right],\label{eq:reducedActionHet}
\end{align}
the gravity multiplet couplings are given by
\begin{align}
    e^{-1} \mathcal{L}_1&= e^{-2 \varphi}\left[\left(R_{\alpha \beta \gamma \delta}\left(\omega_{-}\right)\right)^2-\frac{1}{3} h^{\alpha \beta \gamma} \omega_{3 L}\left(\omega_{-}\right)_{\alpha \beta \gamma}-\frac{1}{2} R^{\alpha \beta \gamma \delta}\left(\omega_{-}\right) F_{\alpha \beta}^{(+) a} F_{\gamma \delta}^{(+) a}\right. \nn\\
    &\left.\qquad\quad\ \,-\frac{1}{8} F_{\alpha \beta}^{(+) a} F_{\beta \gamma}^{(+) a} F_{\gamma \delta}^{(+) b} F_{\delta \alpha}^{(+) b}+\frac{1}{8} F_{\alpha \beta}^{(+) a} F_{\beta \gamma}^{(+) b} F_{\gamma \delta}^{(+) a} F_{\delta \alpha}^{(+) b}\right],
\end{align}
the vector multiplet couplings are given by
\begin{align}
    e^{-1} \mathcal{L}_2&=e^{-2 \varphi}\left[-\frac{1}{8} F_{\alpha \beta}^{(-) a} F_{\gamma \delta}^{(-) a} F_{\alpha \beta}^{(-) b} F_{\gamma \delta}^{(-) b}+\frac{1}{4} F_{\alpha \beta}^{(-) a} F_{\beta \gamma}^{(-) b} F_{\gamma \delta}^{(-) a} F_{\delta \alpha}^{(-) b}\right. \nn\\
    &\qquad\quad\ \,-\frac{1}{8} F_{\alpha \beta}^{(-) a} F_{\beta \gamma}^{(-) a} F_{\gamma \delta}^{(-) b} F_{\delta \alpha}^{(-) b}-\frac{1}{2} P_\gamma^{(-+) a c} P_\gamma^{(-+) b c} F_{\alpha \beta}^{(-) a} F_{\alpha \beta}^{(-) b} \nn\\
    &\qquad\quad\ \,-P_\alpha^{(-+) a b} P_\gamma^{(-+) c b} F_{\alpha \beta}^{(-) c} F_{\beta \gamma}^{(-) a}-P_\alpha^{(-+) a b} P_\gamma^{(-+) c b} F_{\alpha \beta}^{(-) a} F_{\beta \gamma}^{(-) c}\nn \\
    &\qquad\quad\ \,+P_\alpha^{(-+) b c} P_\beta^{(-+) b c} F_{\alpha \gamma}^{(-) a} F_{\gamma \beta}^{(-) a}+2 P_\alpha^{(-+) a b} P_\beta^{(-+) c b} P_\alpha^{(-+) c d} P_\beta^{(-+) a d}\nn \\
    &\qquad\quad\ \,+2 P_\alpha^{(-+) a b} P_\beta^{(-+) c b} P_\beta^{(-+) c d} P_\alpha^{(-+) a d}-2 P_\alpha^{(-+) a b} P_\alpha^{(-+) c b} P_\beta^{(-+) c d} P_\beta^{(-+) a d}\nn \\
    &\qquad\quad\ \,\left.-2 P_\alpha^{(-+) a b} P_\beta^{(-+) a b} P_\alpha^{(-+) c d} P_\beta^{(-+) c d}\right],
\end{align}
and the mixed gravity-vector couplings are given by
\begin{align}
    e^{-1} \mathcal{L}_3&=e^{-2 \varphi} {\left[-\frac{1}{3} h^{\alpha \beta \gamma} \omega_3\left(-Q^{(--)}\right)_{\alpha \beta \gamma}+\frac{1}{8} F_{\alpha \beta}^{(+) a} F_{\gamma \delta}^{(+) a} F_{\alpha \beta}^{(-) b} F_{\gamma \delta}^{(-) a}\right.} \nn\\
    &\qquad\quad\ \, -\frac{1}{4} F_{\alpha \beta}^{(+) a} F_{\beta \gamma}^{(-) b} F_{\gamma \delta}^{(+) a} F_{\delta \alpha}^{(-) b}-\frac{1}{4} F_{\alpha \beta}^{(+) a} F_{\beta \gamma}^{(+) a} F_{\gamma \delta}^{(-) b} F_{\delta \alpha}^{(-) b} \nn\\
    &\qquad\quad\ \, +\frac{1}{2} P_\gamma^{(-+) a b} P_\gamma^{(-+) a c} F_{\alpha \beta}^{(+) b} F_{\alpha \beta}^{(+) c}+P_\alpha^{(-+) b c} P_\gamma^{(-+) b c} F_{\alpha \beta}^{(+) a} F_{\beta \gamma}^{(+) a} \nn\\
    &\qquad\quad\ \, +P_\alpha^{(-+) b a} P_\gamma^{(-+) b c} F_{\alpha \beta}^{(+) a} F_{\beta \gamma}^{(+) c}-P_\alpha^{(-+) b c} P_\gamma^{(-+) b a} F_{\alpha \beta}^{(+) a} F_{\beta \gamma}^{(+) c} \nn\\
    &\qquad\quad\ \, -\frac{1}{2} h^{\alpha \beta \gamma} F_{\alpha \delta}^{(-) a} F_{\beta \gamma}^{(+) b} P_\delta^{(-+) a b}-\frac{1}{2} h^{\alpha \beta \gamma} F_{\alpha \delta}^{(+) b} F_{\beta \gamma}^{(-) a} P_\delta^{(-+) a b}\nn \\
    & \left.\qquad\quad\ \,+h^{\alpha \beta \gamma} F_{\alpha \delta}^{(-) a} F_{\gamma \delta}^{(+) b} P_\beta^{(-+) a b}\right] .
\end{align}
Note also that we have defined
\begin{equation}
    F^{(\pm)\,a}= e^a_i F^{(\pm)\,i},\qquad Q_{\mu ab}^{(\pm\pm)}=e_a^iQ_{\mu ij}^{(\pm\pm)}e_b^j,\qquad P_{\mu ab}^{(\mp\pm)}=e_a^iP_{\mu ij}^{(\mp\pm)}e_b^j.
\end{equation}
Note that, due to lack of worldsheet parity under $B\to -B$ in the original $(d+p)$-dimensional action \eqref{eq:noGaugeFields}, the reduced action \eqref{eq:redAction} is not symmetric under interchange of $O(p-1,1)_-$ and $O(1,p-1)_+$.

\subsection{Consistent Truncations}
One possible consistent truncation of \eqref{eq:redAction} is to set
\begin{equation}
    F^{(+)\,i}=\sqrt{2}\mathcal F^i,\qquad F^{(-)\,i}=0,\qquad g_{ij}=\delta_{ij},\qquad b_{ij}=0,
\end{equation}
which leaves us with the action
\begin{align}
    e^{-1}\mathcal L^{(+)}_{D-d}&=e^{-2\varphi}\left\{R+4(\partial\varphi)^2-\frac{1}{12}\tilde h^2-\frac{1}{4}\qty(\mathcal F^i)^2\nn\right.\\
    &\qquad\qquad\left.+\frac{\alpha'}{8}\qty[R(\omega_-)^2_{\mu\nu\rho\sigma}-R(\omega_-)^{\mu\nu\rho\sigma}\mathcal F^i_{\mu\nu}\mathcal F^i_{\rho\sigma}-\frac{1}{2}\mathcal F^i\mathcal F^i\mathcal F^j\mathcal F^j+\frac{1}{2}\mathcal F^i\mathcal F^j\mathcal F^i\mathcal F^j]\right\},\label{eq:wrongTrunc}
\end{align}
where we have defined
\begin{equation}
    \mathcal F^i=\dd\mathcal A^i,\qquad \mathcal F^i\mathcal F^j\mathcal F^k\mathcal F^l=\mathcal F^i_{\mu}{}^{\nu}\mathcal F^j_{\nu}{}^{\rho}\mathcal F^k_{\rho}{}^{\sigma}\mathcal F^l_{\sigma}{}^{\mu}.
\end{equation}
Note that here, the three-form flux is given by
\begin{equation}
    \tilde h=\dd b-\frac{1}{2}\mathcal A^i\land\mathcal F^i+\frac{\alpha'}{4}\omega_{3L}(\omega_-).\label{eq:hPlusTrunc}
\end{equation}
We will refer to this as the $(+)$-truncation. This was shown to be a consistent truncation by Ref.~\cite{Liu:2023fqq}.

The other possible choice of truncation is to set 
\begin{equation}
    F^{(+)\,i}=0,\qquad F^{(-)\,i}=\sqrt{2}\mathcal F^i,\qquad g_{ij}=\delta_{ij},\qquad b_{ij}=0,\label{eq:secondTrunc}
\end{equation}
which yields the action
\begin{align}
    e^{-1}\mathcal L^{(-)}_{D-d}&=e^{-2\varphi}\left\{R+4(\partial\varphi)^2-\frac{1}{12}\tilde h^2-\frac{1}{4}\qty(\mathcal F^i)^2\right.\nn\\
    &\qquad\qquad\left.+\frac{\alpha'}{8}\qty[R(\omega_-)^2_{\mu\nu\rho\sigma}-\frac{1}{2}\mathcal F^i\mathcal F^i\mathcal F^j\mathcal F^j+\mathcal F^i\mathcal F^j\mathcal F^i\mathcal F^j-\frac{1}{2}\qty(\mathcal F^2_{ij})^2]\right\},\label{eq:rightTrunc}
\end{align}
where
\begin{equation}
    \tilde h=\dd b+\frac{1}{2}\mathcal A^i\land\mathcal F^i+\frac{\alpha'}{4}\omega_{3L}(\omega_-).\label{eq:hMinusTrunc}
\end{equation}
Notice that the sign in front of the gauge Chern-Simons term in~\eqref{eq:hMinusTrunc} is opposite that of~\eqref{eq:hPlusTrunc}. The action~\eqref{eq:rightTrunc} matches what we expect for the heterotic gauge fields, \emph{i.e.}, \eqref{eq:BdR4d}, with an abelian gauge group, so long as we identify the indices $i,j,...$ with $\mathbbm{i},\mathbbm{j},...$. We will refer to this as the $(-)$-truncation.

However, we must check that~\eqref{eq:secondTrunc} is a consistent truncation. This can be seen as follows. First, it is straightforward to verify that this is a consistent truncation at the two-derivative level; therefore, we will focus on the four-derivative part of the action. Since we have already field redefined to the $O(p,p)$ covariant frame, we see that scalar kinetic terms $P^{(-+)}$ and gauge fields $F^{(\pm)}$ always come in $(++)$ and $(--)$ pairs. If we were to look at the tadpoles for the gauge field, which may be thought of as tadpoles for $A^{(+)}$,\footnote{Note that when $g_{ij}$ is non-constant, $F^{(\pm)}$ is not closed, and therefore it does not make sense to refer to $A^{(\pm)}$. It is only when we are discussing tadpoles, when we expand $g_{ij}=\delta_{ij}+\delta g_{ij}$, that it makes sense to refer to $A^{(+)\,i}=A^i+B_i$.} all terms involve a pair of $F^{(+)}$, except for the terms that contain $h$, which contains a gauge Chern-Simons term. However, up to scalar factors that will not affect the $A^{(-)}$ tadpole, the gauge Chern-Simons may be written as
\begin{equation}
    B_i\land F^i= \frac{1}{4}\qty(A^{(+)\,i}\land F^{(+)\,i}-A^{(-)\,i}\land F^{(-)\,i}-2\,\dd\qty(B_i\land A^i))\bigg\vert_{g_{ij}=\delta_{ij},\,b_{ij}=0}.
\end{equation}
Note that the last term will be zero for our purposes later, but more generally can be absorbed into a field redefinition of $b$. The first term is then the only one containing $A^{(+)}$, and again, because of the $O(p-1,1)_-$ symmetry of the action, $A^{(+)}$ appears as a pair, which makes certain that there will be no tadpole if we were to vary the action. 

But we must also be careful about scalar tadpoles. First, note that no scalars appear in $h$. $g_{ij}$ and $b_{ij}$ appear in both $F^{(\pm)}$ and $P^{(-+)}$, but due to the $O(p-1,1)\times O(1,p-1)$ symmetry, there is always at least two of $P^{(-+)}$ and $F^{(+)}$, which ensures a lack of tadpoles, except for the first three terms of $\mathcal L_2$, which are built solely out of $F^{(-)}$. These terms require noting that, when $e_i^a=\delta_i^a+\delta e_i^a$, we have that $\delta e_i^a=-\delta e^i_a$, which implies that
\begin{equation}
    \delta_{g_{ij}} F^{(-)\,a}=\delta e_i^a\qty(F^i+G_i)=0.
\end{equation}
The first three terms of $\mathcal L_2$ may also potentially have $b_{ij}$ tadpoles, but it can be checked that the antisymmetry of $\delta b_{ij}$ kills any possible tadpoles. Note also that $\omega_3(Q^{(--)})$ has no scalar tadpoles since each $Q^{(--)}$ has at least one other $Q^{(--)}$ multiplying it.

Hence, the $(-)$-truncation is indeed consistent. That is to say, this truncation on $T^p$ corresponds precisely to turning on a $U(1)^p$ gauge group for the heterotic gauge fields. This is natural since, after all, heterotic string theory is obtained by compactifying the right movers of the bosonic string on a 16-dimensional torus.

It is worth noting that for the two truncations $F^i=\pm G_i$, we find that~\eqref{eq:secondFieldsRedefs} yields $\delta A^i=\pm \delta B_i$, which is to say that the second set of field redefinitions~\eqref{eq:secondFieldsRedefs} commutes with both truncations. Note also that $\mathcal L^{(+)}$ and $\mathcal L^{(-)}$ match the two $O(d+p,d)$-invariant actions found from double field theory in~\cite{Baron:2017dvb}, where it was remarked that the $(-)$-truncation matched the BdR action and the $(+)$-truncation action seemingly had no connection to string theory. However, we see that the $(+)$-truncation is indeed realized as a particular truncation of string theory.

There is, of course, a third possible truncation that truncates everything. It is straightforward to see that 
\begin{equation}
    F^{(+)\,i}=0,\qquad  F^{(-)\,i}=0,\qquad g_{ij}=\delta_{ij},\qquad b_{ij}=0,
\end{equation}
is a consistent truncation, and will leave us with just the heterotic action in $d$ dimensions. We will refer to this as the $(0)$-truncation for convenience.

It should be noted that we may ``mix and match'' truncations, in the sense that we can write $p=p_\star+p_++p_-+p_0$ and apply the $(+)$-truncation along $T^{p_+}$, apply the $(-)$-truncation along $T^{p_-}$, apply the $(0)$-truncation along $T^{p_0}$, and apply no truncation along $T^{p_\star}$. Since the torus directions do not talk to each other, this is guaranteed to be consistent. We will eventually be interested in the two cases given by $p_\pm=1$, $p_{\mp}=0$, $p_\star=0$, and $p_0=5$ for ten-dimensional heterotic supergravity.

\subsection{Embedding $O(d+p,d)$ in $O(d+p,d+p)$}\label{sec:OdpdfromOdd}
We now discuss how $O(d+p,d)$ arises as a diagonal subgroup of $O(d+p,d+p)$. Here, we closely follow the derivation in~\cite{Hu:2025aji}, which found the embedding for the $(+)$-truncation. However, we now have two truncations that we are interested in embedding, and we will find the explicit isomorphism between the two $O(d+p,d)$ subgroups. The starting point is to suppose we begin in $D=\mathfrak d+d+p$ dimensions and reduce on $T^{d+p}$ to $\mathfrak d$ dimensions. We let the $T^{d+p}$ index be $\hat i=\{i,\mathfrak a\}$, where $i$ is an index on $T^d$ and $\mathfrak a$ is an index on $T^p$. If we first perform a $T^p$ reduction followed by a $T^d$ reduction, the fields decompose as
\begin{align}
    \dd s^2_D&=g_{\mu\nu}\dd x^\mu\dd x^\nu+g_{\hat i\hat j}\eta^{\hat i}\eta^{\hat j},\qquad \eta^{\hat i}=\dd y^{\hat i}+\mathcal A_\mu^{\hat i}\dd x^\mu,\nn\\
    B&=\frac{1}{2}b_{\mu\nu}\dd x^\mu\land\dd x^\nu+B_{\mu\hat i}\dd x^\mu\land\eta^{\hat i}+\frac{1}{2}b_{\hat i\hat j}\eta^{\hat i}\land\eta^{\hat j},
\end{align}
where, if we perform the $(\pm)$-truncation along $T^p$,
\begin{align}
    g_{\hat i\hat j}&=\begin{pmatrix}
        g_{ij}+\frac{1}{2}\mathcal A_i^{\mathfrak a}\mathcal A_j^{\mathfrak a}&\quad \frac{1}{\sqrt{2}}\mathcal A_i^{\mathfrak b}\\ \frac{1}{\sqrt{2}}\mathcal A_j^{\mathfrak a}&\quad\delta_{\mathfrak a\mathfrak b}
    \end{pmatrix},\qquad A_\mu^{\hat i}=\begin{pmatrix}
        A_\mu^i\\ \frac{1}{\sqrt{2}}\mathcal A_\mu^{\mathfrak a}
    \end{pmatrix},\nn\\
    b_{\hat i\hat j}&=\begin{pmatrix}
        b_{ij}&\quad\pm\frac{1}{\sqrt{2}}\mathcal A_i^{\mathfrak b}\\ \mp\frac{1}{\sqrt{2}}\mathcal A_j^{\mathfrak a}&\quad 0
    \end{pmatrix},\qquad B_{\mu\hat i}=\begin{pmatrix}
        B_{\mu i}\pm\frac{1}{2}\mathcal A_\mu^{\mathfrak a} \mathcal A_i^{\mathfrak a}\\
        \pm\frac{1}{\sqrt{2}}\mathcal A_\mu^{\mathfrak a}
    \end{pmatrix}.\label{eq:truncFieldExp}
\end{align}
Note that in~\eqref{eq:truncFieldExp}, the upper sign corresponds to the $(+)$-truncation and the lower sign corresponds to the $(-)$-truncation. Here $x^\mu$ are the $\mathfrak d$-dimensional coordinates, while $y^{\hat i}$ are the coordinates on $T^{d+p}$.

The action without truncation naturally has an $O(d+p,d+p)$ invariance, and can be written in terms of the generalized metric
\begin{equation}
    \mathcal H=\begin{pmatrix}
        g_{\hat i\hat j}-b_{\hat i\hat k}g^{\hat k\hat l}b_{\hat l\hat j} & \quad b_{\hat i\hat k}g^{\hat k\hat j}\\
        -g^{\hat i\hat k}b_{\hat k\hat j} & \quad g^{\hat i\hat j}
    \end{pmatrix},
\end{equation}
the generalized gauge field
\begin{equation}
    \mathbb A_\mu=\begin{pmatrix}
        A^{\hat i}_\mu\\B_{\mu\hat i}
    \end{pmatrix},
\end{equation}
with field strength $\mathbb F=\dd\mathbb A$, and the $O(d+p,d+p)$-invariant bilinear form
\begin{equation}
    \eta = \begin{pmatrix}
        0&\quad\openone\\
        \openone&\quad0
    \end{pmatrix}.
\end{equation}
Under an $O(d+p,d+p)$ transformation, these transform as
\begin{equation}
    \mathcal H\to(\Omega^{-1})^T\mathcal H\Omega^{-1},\qquad \mathbb{A}\to \Omega\mathbb A,\qquad \Omega^T\eta\Omega=\eta.
\end{equation}
The last equation is simply the statement that $\Omega\in O(d+p,d+p)$. Note that $\mathcal H$ satisfies
\begin{equation}
    \eta\mathcal H\eta = \mathcal H^{-1},
\end{equation}
which implies that the generalized metric $\mathcal H\in O(d+p,d+p)$ parametrizes the scalar coset $O(d+p,d+p)/(O(d+p-1,1)\times O(1,d+p-1))$.

Plugging in~\eqref{eq:truncFieldExp} gives
\begin{align}
    \mathcal H^{(\pm)}&=\begin{pmatrix}
        \quad g_{ij}+c^{(\pm)}_{ki}g^{kl}c^{(\pm)}_{lj}+\mathcal A_i^{\mathfrak a}\mathcal A_j^{\mathfrak a}&\quad \frac{1}{\sqrt{2}}(c^{(\pm)}_{ki}g^{kl}\mathcal A_l^{\mathfrak b}+\mathcal A_i^{\mathfrak b})&\quad \mp g^{jk}c^{(\pm)}_{ki}&\quad \pm\frac{1}{\sqrt{2}}(c^{(\pm)}_{ki}g^{kl}\mathcal A_l^{\mathfrak b}+\mathcal A_i^{\mathfrak b})\\
        \frac{1}{\sqrt{2}}(c^{(\pm)}_{kj}g^{kl}\mathcal A_l^{\mathfrak a}+\mathcal A_j^{\mathfrak a})&\quad \delta^{\mathfrak a\mathfrak b}+\frac{1}{2}\mathcal A_k^{\mathfrak a}g^{kl}\mathcal A^{\mathfrak b}_l&\quad\mp\frac{1}{\sqrt{2}}g^{jk}\mathcal A_k^{\mathfrak a}&\quad \pm\frac{1}{2}\mathcal A_k^{\mathfrak a}g^{kl}\mathcal A^{\mathfrak b}_l\\
        \mp g^{ik}c^{(\pm)}_{kj}&\quad\mp\frac{1}{\sqrt{2}}g^{ik}\mathcal A_k^{\mathfrak b}&g^{ij}&\quad -\frac{1}{\sqrt{2}}g^{ik}\mathcal A_k^{\mathfrak b}\\
        \pm\frac{1}{\sqrt{2}}(c^{(\pm)}_{kj}g^{kl}\mathcal A_l^{\mathfrak a}+\mathcal A_j^{\mathfrak a})&\quad\pm\frac{1}{2}\mathcal A_k^{\mathfrak a}g^{kl}\mathcal A^{\mathfrak b}_l&\quad -\frac{1}{\sqrt{2}}g^{jk}\mathcal A_k^{\mathfrak a}&\quad \delta_{\mathfrak a\mathfrak b}+\frac{1}{2}\mathcal A_k^{\mathfrak a}g^{kl}\mathcal A^{\mathfrak b}_l
    \end{pmatrix},\nn\\
    \mathbb A^{(\pm)}_\mu&=\begin{pmatrix}
        A_\mu^i\\ \frac{1}{\sqrt{2}}\mathcal A_\mu^{\mathfrak a}\\B_{\mu i}\pm\frac{1}{2}\mathcal A_\mu^{\mathfrak a}\mathcal A_i^{\mathfrak a}\\
        \pm\frac{1}{\sqrt{2}}\mathcal A_\mu^{\mathfrak a}
    \end{pmatrix},\label{eq:HbeforeRot}
\end{align}
where
\begin{equation}
    c^{(\pm)}_{ij}=b_{ij}\pm\frac{1}{2}\mathcal A_i^{\mathfrak a}\mathcal A_j^{\mathfrak a}.
\end{equation}
Let $\iota$ denote the $B$-inversion map $b_{ij}\to-b_{ij}$, $B_{\mu i}\to-B_{\mu i}$. Then we see that
\begin{equation}
    \mathcal H^{(\pm)}=U^T\iota\qty(\mathcal H^{(\mp)})U,\qquad \mathbb A^{(\pm)}=U^T\iota\qty(\mathbb A^{(\mp)}),\label{eq:pmCosetRelations}
\end{equation}
where
\begin{equation}
    U=\begin{pmatrix}
        \openone&\quad0\\0&\quad-\openone
    \end{pmatrix}.\label{eq:Udef}
\end{equation}
Note that when $b_{ij}=0$ and $B_{\mu i}=0$, this is simply a linear transformation. 

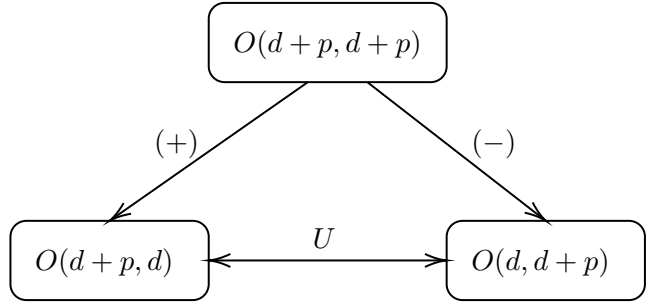
\begin{figure}
    \centering
    \tikzset{every picture/.style={line width=0.75pt}} 
    
    \begin{tikzpicture}[x=0.75pt,y=0.75pt,yscale=-1,xscale=1]
    
    \draw   (240,118) .. controls (240,113.58) and (243.58,110) .. (248,110) -- (352,110) .. controls (356.42,110) and (360,113.58) .. (360,118) -- (360,142) .. controls (360,146.42) and (356.42,150) .. (352,150) -- (248,150) .. controls (243.58,150) and (240,146.42) .. (240,142) -- cycle ;
    \draw   (140,228) .. controls (140,223.58) and (143.58,220) .. (148,220) -- (232,220) .. controls (236.42,220) and (240,223.58) .. (240,228) -- (240,252) .. controls (240,256.42) and (236.42,260) .. (232,260) -- (148,260) .. controls (143.58,260) and (140,256.42) .. (140,252) -- cycle ;
    \draw   (360,228) .. controls (360,223.58) and (363.58,220) .. (368,220) -- (452,220) .. controls (456.42,220) and (460,223.58) .. (460,228) -- (460,252) .. controls (460,256.42) and (456.42,260) .. (452,260) -- (368,260) .. controls (363.58,260) and (360,256.42) .. (360,252) -- cycle ;
    \draw    (290,150) -- (191.64,218.85) ;
    \draw [shift={(190,220)}, rotate = 325.01] [color={rgb, 255:red, 0; green, 0; blue, 0 }  ][line width=0.75]    (10.93,-3.29) .. controls (6.95,-1.4) and (3.31,-0.3) .. (0,0) .. controls (3.31,0.3) and (6.95,1.4) .. (10.93,3.29)   ;
    \draw    (320,150) -- (408.42,218.77) ;
    \draw [shift={(410,220)}, rotate = 217.87] [color={rgb, 255:red, 0; green, 0; blue, 0 }  ][line width=0.75]    (10.93,-3.29) .. controls (6.95,-1.4) and (3.31,-0.3) .. (0,0) .. controls (3.31,0.3) and (6.95,1.4) .. (10.93,3.29)   ;
    \draw    (242,240) -- (358,240) ;
    \draw [shift={(360,240)}, rotate = 180] [color={rgb, 255:red, 0; green, 0; blue, 0 }  ][line width=0.75]    (10.93,-3.29) .. controls (6.95,-1.4) and (3.31,-0.3) .. (0,0) .. controls (3.31,0.3) and (6.95,1.4) .. (10.93,3.29)   ;
    \draw [shift={(240,240)}, rotate = 0] [color={rgb, 255:red, 0; green, 0; blue, 0 }  ][line width=0.75]    (10.93,-3.29) .. controls (6.95,-1.4) and (3.31,-0.3) .. (0,0) .. controls (3.31,0.3) and (6.95,1.4) .. (10.93,3.29)   ;
    
    \draw (251,122.4) node [anchor=north west][inner sep=0.75pt]    {$O( d+p,d+p)$};
    \draw (151,232.4) node [anchor=north west][inner sep=0.75pt]    {$O( d+p,d)$};
    \draw (371,232.4) node [anchor=north west][inner sep=0.75pt]    {$O( d,d+p)$};
    \draw (211,172.4) node [anchor=north west][inner sep=0.75pt]    {$( +)$};
    \draw (371,172.4) node [anchor=north west][inner sep=0.75pt]    {$( -)$};
    \draw (291,222.4) node [anchor=north west][inner sep=0.75pt]    {$U$};

    \end{tikzpicture}
    \caption{The $(+)$- and $(-)$-truncations restrict us to two $O(d+p,d)$ subgroups of $O(d+p,d+p)$, while a similarity transformation with $U$ provides an isomorphism between them.}
    \label{fig:OddIsomorphism}
\end{figure}

The relation~\eqref{eq:pmCosetRelations} is quite useful. Suppose we have a transformation $\Omega\in O(d+p,d+p)$ that respects the $(\pm)$-truncation, in the sense that it keeps $A^i=\pm B_i$. $\Omega$ and $U$ are constant matrices and therefore do not depend on $b$. Hence, we can move them past $\iota$.
\begin{equation}
    (\Omega^{-1})^T\mathcal H^{(\pm)}\Omega^{-1}=\iota\qty((\Omega^{-1})^TU^T\mathcal H^{(\mp)}U\Omega^{-1}),\qquad \Omega\mathbb A^{(\pm)}=\iota\qty(\Omega U^T\mathbb A^{(\mp)}).
\end{equation}
So we see that $\Omega U$ is a transformation respecting the $(\mp)$-truncation. However, $U^T\eta U=-\eta$, which means that
\begin{equation}
    U^T\Omega^T\eta\Omega U=-\eta,
\end{equation}
so our transformation lies outside $O(d+p,d+p)$. We can fix this by conjugating again with $U$. Therefore, we see that if $\Omega\in O(d+p,d+p)$ preserves the $(\pm)$-truncation, then $U^T\Omega U\in O(d+p,d+p)$ preserves the $(\mp)$-truncation. This is thus an isomorphism between the two copies of $O(d+p,d)$, as illustrated in Figure~\ref{fig:OddIsomorphism}.

Now, we need to make the $O(d+p,d)$ subgroups manifest, so we perform a change of basis
\begin{equation}
    \mathcal H^{(\pm)}\to V\mathcal H^{(\pm)} V^T,\qquad \mathbb A^{(\pm)}\to V\mathbb A^{(\pm)},\qquad \eta\to V\eta V^T,\qquad V=\begin{pmatrix}
        \openone&\quad0&\quad0&\quad0\\
        0&\quad\frac{1}{\sqrt{2}}\openone&\quad 0&\mp\frac{1}{\sqrt{2}}\openone\\
        0&\quad0&\quad\openone&\quad 0\\
        0&\quad\frac{1}{\sqrt{2}}\openone&\quad0&\quad\pm\frac{1}{\sqrt{2}}\openone
    \end{pmatrix}.\label{eq:VtransH}
\end{equation}
Note that this change of basis depends on the choice of truncation.

We then find that
\begin{align}
    \mathcal H^{(\pm)}&=\begin{pmatrix}
        \quad g_{ij}+c^{(\pm)}_{ki}g^{kl}c^{(\pm)}_{lj}+\mathcal A_i^{\mathfrak a}\mathcal A_j^{\mathfrak a}&\quad 0&\quad \mp g^{jk}c^{(\pm)}_{ki}&\quad \pm c^{(\pm)}_{ki}g^{kl}\mathcal A_l^{\mathfrak b}\pm \mathcal A_i^{\mathfrak b}\\
        0&\quad \delta^{\mathfrak a\mathfrak b}&\quad0&\quad 0\\
        \mp g^{ik}c^{(\pm)}_{kj}&\quad0&g^{ij}&\quad -g^{ik}\mathcal A_k^{\mathfrak b}\\
        \pm c^{(\pm)}_{kj}g^{kl}\mathcal A_l^{\mathfrak a}\pm \mathcal A_j^{\mathfrak a}&\quad0&\quad -g^{jk}\mathcal A_k^{\mathfrak a}&\quad \delta_{\mathfrak a\mathfrak b}+\mathcal A_k^{\mathfrak a}g^{kl}\mathcal A^{\mathfrak b}_l
    \end{pmatrix},\nn\\
    \mathbb A^{(\pm)}_\mu&=\begin{pmatrix}
        A_\mu^i\\ 0\\B_{\mu i}\pm\frac{1}{2}\mathcal A_\mu^{\mathfrak a}\mathcal A_i^{\mathfrak a}\\
        \mathcal A_\mu^{\mathfrak a}
    \end{pmatrix},\qquad\eta=\begin{pmatrix}
        0&\quad0&\quad\delta_i{}^{j}&\quad0\\
        0&\quad\mp\delta^{\mathfrak a\mathfrak b}&\quad0&\quad0\\
        \delta^{i}{}_j&\quad0&\quad0&\quad0\\
        0&\quad0&\quad0&\quad\pm\delta_{\mathfrak a\mathfrak b}
    \end{pmatrix}.\label{eq:HafterRot}
\end{align}
It is then clear that we get an $O(d+p,d)$ coset by erasing the second row/column:
\begin{align}
    \bar{\mathcal H}^{(\pm)}&=\begin{pmatrix}
        \quad g_{ij}+c^{(\pm)}_{ki}g^{kl}c^{(\pm)}_{lj}+\mathcal A_i^{\mathfrak a}\mathcal A_j^{\mathfrak a}&\quad \mp g^{jk}c^{(\pm)}_{ki}&\quad \pm c^{(\pm)}_{ki}g^{kl}\mathcal A_l^{\mathfrak b}\pm\mathcal A_i^{\mathfrak b}\\
        \mp g^{ik}c^{(\pm)}_{kj}&g^{ij}&\quad -g^{ik}\mathcal A_k^{\mathfrak b}\\
        \pm c^{(\pm)}_{kj}g^{kl}\mathcal A_l^{\mathfrak a}\pm\mathcal A_j^{\mathfrak a}&\quad -g^{jk}\mathcal A_k^{\mathfrak a}&\quad \delta_{\mathfrak a\mathfrak b}+\mathcal A_k^{\mathfrak a}g^{kl}\mathcal A^{\mathfrak a}_l
    \end{pmatrix},\nn\\
    \bar{\mathbb A}^{(\pm)}_\mu&=\begin{pmatrix}
        A_\mu^i\\B_{\mu i}\pm\frac{1}{2}\mathcal A_\mu^{\mathfrak a}\mathcal A_i^{\mathfrak a}\\
        \mathcal A_\mu^{\mathfrak a}
    \end{pmatrix},\qquad
    \bar\eta=\begin{pmatrix}
        0&\quad\delta_i{}^{j}&\quad0\\
        \delta^i{}_{j}&\quad0&\quad0\\
        0&\quad0&\quad\pm\delta_{\mathfrak a\mathfrak b}
    \end{pmatrix}.\label{eq:OdpdCoset}
\end{align}
For the $(+)$-truncation, this matches what we expect~\cite{Maharana:1992my,Sen:1994fa,Hohm:2011ex}, and was found previously in~\cite{Hu:2025aji}. For the $(-)$-truncation, we must do some further rearrangement. We define
\begin{equation}
    \bar U=\begin{pmatrix}
        \openone&\quad0&\quad0\\0&\quad-\openone&\quad0\\0&\quad0&\quad-\openone
    \end{pmatrix},
\end{equation}
analogous to the definition of $U$ in~\eqref{eq:Udef}. We then do a change of basis
\begin{equation}
    \mathcal H^{(-)}\to \bar U^T\iota\qty(\mathcal H^{(-)}) \bar U,\qquad \mathbb A^{(-)}\to \bar U\iota\qty(\mathbb A^{(-)}),\qquad \eta\to -\bar U^T\eta \bar U.
\end{equation}
Note that the minus sign flip in $\bar\eta$ is perfectly legal, as flipping the sign of $\bar\eta$ is equivalent to a choice of sign convention. This also involves a sign flip $\iota$ of $b$ and $B_i$, which corresponds to a choice of worldsheet parity. In this basis, we recover the expected coset definition~\cite{Maharana:1992my,Sen:1994fa,Hohm:2011ex}. Equivalently, this gives an isomorphism between the copies of $O(d+p,d)$.

\section{The 4D Theory}\label{sec:4D}
From this point forward, we will confine our attention to four dimensions. In particular, we will be interested in generating four-derivative corrections to the Kerr-Sen solution. At the two-derivative level, this just means we are interested in solutions to the four-dimensional bosonic action
\begin{equation}
    e^{-1}\mathcal L_{2\partial}=e^{-2\phi}\qty[R+4(\partial\phi)^2-\frac{1}{12}H^2-\frac{1}{4}\mathcal F^2],\qquad H=\dd B-\frac{1}{2}\mathcal A\land \mathcal F,\qquad \mathcal F=\dd\mathcal A.
\end{equation}
This action can be obtained by reducing heterotic supergravity (without gauge fields) on $T^6$ and taking the $(\pm)$-truncation along one of the six torus directions and the $(0)$-truncation along the other five. Note that at the two-derivative level, we can always flip the sign of the gauge Chern-Simons term by flipping $B\to -B$. That is to say, the $(+)$- and $(-)$-truncations differ only by a choice of sign for $B$ at the two-derivative level. From Section~\ref{sec:OdpdfromOdd}, we know that this action naturally has an $O(2,1)$ symmetry. Moreover, as shown in Appendix~\ref{app:susy}, the $(-)$-truncation preserves $\mathcal N=1$ supersymmetry, whereas the $(+)$-truncation (with five of the six gauge fields truncated away) does not seem to lead to a supersymmetric theory. However, at the two-derivative level, the two nevertheless appear identical in the bosonic sector.

The two-derivative Kerr-Sen solution was first obtained in Ref.~\cite{Sen:1992ua}. The starting point is to embed the Kerr solution into heterotic supergravity as
\begin{align}
    \dd s^2&=-\left(1-\frac{2\mu r}{\Sigma}\right)\dd t^2-\frac{4\mu ar(1-x^2)}{\Sigma}\,\dd t\,\dd y+\Sigma\left(\frac{\dd r^2}{\Delta}+\frac{\dd x^2}{1-x^2}\right)\nn\\
    &\quad+\qty(r^2+a^2+\frac{2\mu ra^2(1-x^2)}{\Sigma})(1-x^2)\,\dd y^2,\nn\\
    \phi&=0,\qquad B=0,\qquad \mathcal A=0,\label{eq:Kerr}
\end{align}
where
\begin{equation}
    \Sigma=r^2+a^2x^2,\qquad \Delta=r^2-2\mu r+a^2.
\end{equation}
One then reduces along the time direction, assembles the resulting fields into $O(2,1)$ cosets as in~\eqref{eq:OdpdCoset}, and applies a particular $O(2,1)$ transformation
\begin{equation}
    \Omega^{(0)}=\bar V\,\bar\Omega\,\bar V^T,
\end{equation}
where
\begin{equation}
    \bar\Omega=\begin{pmatrix}
        \cosh\beta &\quad \sinh\beta &\quad 0 \\
 \sinh\beta &\quad \cosh\beta &\quad 0 \\
 0 &\quad 0 &\quad 1
    \end{pmatrix},\qquad \bar V=\frac{1}{\sqrt{2}}\begin{pmatrix}
        1 &\quad -1&\quad0 \\
 1&\quad 1&\quad0\\
 0&\quad0&\quad\sqrt2
    \end{pmatrix},\label{eq:KStwoDerTrans}
\end{equation}
to obtain a solution
\begin{align}
    \dd s^{\prime2}&=-\frac{\Sigma\tilde\Delta}{\Upsilon^2}\dd t^2-\frac{4 a \mu r \left(1-x^2\right) \cosh ^2\frac{\beta }{2} \Sigma}{\Upsilon^2}\dd t\,\dd y+\Sigma\qty(\frac{\dd r^2}{\Delta}+\frac{\dd x^2}{1-x^2})\nn\\
    &\quad+\frac{(1-x^2)\Sigma}{\Upsilon^2}\qty(\left(r^2+a^2\right) \Sigma +2 \mu r a^2 \left(1-x^2\right)+4 \mu r \left(r^2+a^2\right) \sinh ^2\frac{\beta }{2}+4 \mu^2 r^2 \sinh
   ^4\frac{\beta }{2})\dd y^2,\nn\\
   \phi'&=-\frac{1}{2}\ln\frac{\Upsilon}{\Sigma},\qquad B'=\frac{2\mu a  r}{\Upsilon}(1-x^2)\sinh^2\frac{\beta}{2}\dd t\wedge\dd y,\nn\\
   \mathcal A'&=\frac{\sqrt{2}\mu r\sinh\beta}{\Upsilon}\qty(\dd t-a (1-x^2)\dd y),\label{eq:KerrSen2der}
\end{align}
where we have defined
\begin{align}
    \tilde\Delta&=r^2+a^2x^2-2\mu r,\nn\\
    \Upsilon&=r^2+a^2 x^2+2\mu r \sinh^2\frac{\beta}{2},\nn\\
    \Xi&=  \left(r^2+a^2x^2-\mu r\right)\cosh \beta+\mu r.
\end{align}

Note that this solution is non-extremal and therefore not BPS, although the extremal limit may potentially be BPS. In particular, the static limit, which corresponds to the Gibbons-Maeda-Garfinkle-Horowitz-Strominger (GMGHS) solution~\cite{Gibbons:1982ih,Gibbons:1987ps,Garfinkle:1990qj}, is supersymmetric in the extremal limit~\cite{Kallosh:1992ii}. Nevertheless, it is known that the extremal Kerr solution is not supersymmetric~\cite{Bardeen:1999px}. In the context of Einstein-Maxwell theory, the Kerr-Newman black hole is supersymmetric only for the case of zero angular momentum~\cite{Gibbons:1982fy,Tod:1983pm}, which hints that we should expect something similar for the Kerr-Sen case. For the heterotic theory, we expect the supersymmetry bound in flat space to be
\begin{equation}
    M\ge \sqrt{2}|Q|,
\end{equation}
whereas the extremality bound for Kerr-Sen black holes is given by~\cite{Sen:1992ua}
\begin{equation}
    M^2\ge 2Q^2+|J|.
\end{equation}
The only way to saturate both inequalities is to have $J=0$, which means that we should only expect the extremal GMGHS black hole to be BPS.

We now wish to perform this boost at the four-derivative level for both choices of truncation.

\subsection{Four-derivative Kerr-Sen Solutions}
To begin, we need a four-derivative seed solution. Since the Kerr solution has no gauge fields (or, equivalently, is a solution to the $(0)$-truncation), the four-derivative seed solution is the same for both the $(+)$- and $(-)$-truncations. Unfortunately, no closed-form analytic expression is known for the four-derivative corrections to the Kerr solution in heterotic supergravity. Instead, corrections must be computed perturbatively in the spin parameter~\cite{Pani:2009wy,Yunes:2009hc,Konno:2009kg,Yagi:2012ya,Mignemi:1992pm,Cano:2019ore}
\begin{equation}
    \chi=\frac{a}{\mu}.
\end{equation}
The corrections were originally found in~\cite{Cano:2019ore}. The metric remains uncorrected in the Einstein frame, and hence receives only corrections from the dilaton in the string frame. The gauge field $\mathcal A$ remains zero since the $(0)$-truncation is a consistent truncation, while the remaining fields receive corrections
\begin{align}
    B&=\alpha'\Lambda\,\dd t\land\dd y,\nn\\
    \Lambda&=\qty(\frac{x^2   \left(-18 \mu ^2+5 r^2+5 \mu  r\right)}{16 r^3}-\frac{ \left(-2 \mu ^2+5 r^2+5 \mu  r\right)}{16 r^3})\chi\nn\\
    &\quad-\left\{\frac{x^2  }{160 r^5}\Big[5 r^4+5 \mu  r^3+4 \mu ^2 r^2 \left(15 x^2-14\right)+2 \mu ^3 r \left(90 x^2-89\right)\right.\nn\\
    &\qquad\quad\left.+80 \mu ^4 \left(1-5 x^2\right)\Big]+\frac{ \left(2 \mu ^3+5 r^3+5 \mu  r^2+4 \mu ^2 r\right)}{160 r^4} \right\}\chi^3\nn\\
    &\quad+\left\{\frac{x^2  }{8960 r^7}\Big[-105 r^6-105 \mu  r^5+10 \mu ^2 r^4 \left(10 x^2-19\right)+60 \mu ^3 r^3 \left(5 x^2-6\right)\right.\nn\\
    &\qquad\quad+8 \mu ^4 r^2 \left(455 x^4-395 x^2-63\right)+200 \mu ^5 r \left(91 x^4-89 x^2-2\right)+1680 \mu ^6 x^2 \left(5-21 x^2\right)\Big]\nn\\
    &\qquad\quad\left.+\frac{3  \left(8 \mu ^4+35 r^4+35 \mu  r^3+30 \mu ^2 r^2+20 \mu ^3 r\right)}{8960 r^5}\right\}\chi^5+\mathcal O(\chi^7),\nn\\
    \phi&=\alpha'\left[-\frac{\mu }{6 r^3}-\frac{1}{8 r^2}-\frac{1}{8 \mu  r}+ \chi^2\left(\frac{\mu ^2 }{80 r^4}+\frac{\mu }{40 r^3}+\frac{1}{32 r^2}+\frac{1}{32 \mu  r}\right)\right.\nn\\
    &\qquad+\chi ^2 x^2 \left(\frac{6 \mu ^3 }{5 r^5}+\frac{21 \mu ^2 }{40 r^4}+\frac{7 \mu }{40 r^3}\right)+\chi ^4 \left(\frac{\mu ^3 }{280 r^5}+\frac{\mu ^2 }{112 r^4}+\frac{3 \mu  }{224 r^3}+\frac{1}{64 r^2}+\frac{1}{64 \mu  r}\right)\nn\\
    &\qquad\left.+\chi ^4 x^4 \left(-\frac{45 \mu ^5}{14 r^7}-\frac{55 \mu ^4 }{56 r^6}-\frac{11 \mu ^3 }{56 r^5}\right)+\chi ^4 x^2 \left(-\frac{\mu ^4}{28 r^6}-\frac{3 \mu ^3}{70 r^5}-\frac{3 \mu ^2 }{112 r^4}-\frac{\mu }{112 r^3}\right)+\mathcal O(\chi^6)\right].\label{eq:KerrExpansion}
\end{align}
Here, we have only presented the first few orders of $\chi$ for the sake of space.

The procedure of~\cite{Sen:1992ua} works quite smoothly at the two-derivative level since the action reduced along the time ``circle'' is automatically $O(2,1)$ invariant. At the four-derivative level, that invariance appears only after appropriate field redefinitions. In particular, the minimal set of such field redefinitions was found by Ref.~\cite{Jayaprakash:2024xlr}; however, the authors truncated the heterotic gauge fields from the beginning. Of course, we may uplift to five dimensions as a trick to reinstate the gauge fields, as discussed in Section~\ref{sec:OdpdfromOdd}.

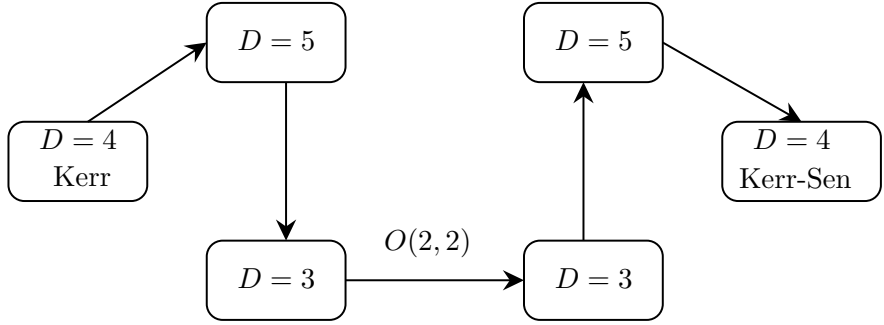
\begin{figure}
    \centering

\tikzset{every picture/.style={line width=0.75pt}} 

\begin{tikzpicture}[x=0.75pt,y=0.75pt,yscale=-1,xscale=1]

\draw   (40,118) .. controls (40,113.58) and (43.58,110) .. (48,110) -- (102,110) .. controls (106.42,110) and (110,113.58) .. (110,118) -- (110,142) .. controls (110,146.42) and (106.42,150) .. (102,150) -- (48,150) .. controls (43.58,150) and (40,146.42) .. (40,142) -- cycle ;
\draw   (140,58) .. controls (140,53.58) and (143.58,50) .. (148,50) -- (202,50) .. controls (206.42,50) and (210,53.58) .. (210,58) -- (210,82) .. controls (210,86.42) and (206.42,90) .. (202,90) -- (148,90) .. controls (143.58,90) and (140,86.42) .. (140,82) -- cycle ;
\draw   (140,178) .. controls (140,173.58) and (143.58,170) .. (148,170) -- (202,170) .. controls (206.42,170) and (210,173.58) .. (210,178) -- (210,202) .. controls (210,206.42) and (206.42,210) .. (202,210) -- (148,210) .. controls (143.58,210) and (140,206.42) .. (140,202) -- cycle ;
\draw   (300,58) .. controls (300,53.58) and (303.58,50) .. (308,50) -- (362,50) .. controls (366.42,50) and (370,53.58) .. (370,58) -- (370,82) .. controls (370,86.42) and (366.42,90) .. (362,90) -- (308,90) .. controls (303.58,90) and (300,86.42) .. (300,82) -- cycle ;
\draw   (300,178) .. controls (300,173.58) and (303.58,170) .. (308,170) -- (362,170) .. controls (366.42,170) and (370,173.58) .. (370,178) -- (370,202) .. controls (370,206.42) and (366.42,210) .. (362,210) -- (308,210) .. controls (303.58,210) and (300,206.42) .. (300,202) -- cycle ;
\draw   (400,118) .. controls (400,113.58) and (403.58,110) .. (408,110) -- (472,110) .. controls (476.42,110) and (480,113.58) .. (480,118) -- (480,142) .. controls (480,146.42) and (476.42,150) .. (472,150) -- (408,150) .. controls (403.58,150) and (400,146.42) .. (400,142) -- cycle ;
\draw [color={rgb, 255:red, 0; green, 0; blue, 0 }  ,draw opacity=1 ]   (80,110) -- (137.5,71.66) ;
\draw [shift={(140,70)}, rotate = 146.31] [fill={rgb, 255:red, 0; green, 0; blue, 0 }  ,fill opacity=1 ][line width=0.08]  [draw opacity=0] (10.72,-5.15) -- (0,0) -- (10.72,5.15) -- (7.12,0) -- cycle    ;
\draw [color={rgb, 255:red, 0; green, 0; blue, 0 }  ,draw opacity=1 ]   (180,90) -- (180,167) ;
\draw [shift={(180,170)}, rotate = 270] [fill={rgb, 255:red, 0; green, 0; blue, 0 }  ,fill opacity=1 ][line width=0.08]  [draw opacity=0] (10.72,-5.15) -- (0,0) -- (10.72,5.15) -- (7.12,0) -- cycle    ;
\draw [color={rgb, 255:red, 0; green, 0; blue, 0 }  ,draw opacity=1 ]   (210,190) -- (297,190) ;
\draw [shift={(300,190)}, rotate = 180] [fill={rgb, 255:red, 0; green, 0; blue, 0 }  ,fill opacity=1 ][line width=0.08]  [draw opacity=0] (10.72,-5.15) -- (0,0) -- (10.72,5.15) -- (7.12,0) -- cycle    ;
\draw [color={rgb, 255:red, 0; green, 0; blue, 0 }  ,draw opacity=1 ]   (330,170) -- (330,93) ;
\draw [shift={(330,90)}, rotate = 90] [fill={rgb, 255:red, 0; green, 0; blue, 0 }  ,fill opacity=1 ][line width=0.08]  [draw opacity=0] (10.72,-5.15) -- (0,0) -- (10.72,5.15) -- (7.12,0) -- cycle    ;
\draw [color={rgb, 255:red, 0; green, 0; blue, 0 }  ,draw opacity=1 ]   (370,70) -- (437.4,108.51) ;
\draw [shift={(440,110)}, rotate = 209.74] [fill={rgb, 255:red, 0; green, 0; blue, 0 }  ,fill opacity=1 ][line width=0.08]  [draw opacity=0] (10.72,-5.15) -- (0,0) -- (10.72,5.15) -- (7.12,0) -- cycle    ;

\draw (61,131) node [anchor=north west][inner sep=0.75pt]   [align=left] {Kerr};
\draw (54,112.4) node [anchor=north west][inner sep=0.75pt]    {$D=4$};
\draw (154,62.4) node [anchor=north west][inner sep=0.75pt]    {$D=5$};
\draw (154,182.4) node [anchor=north west][inner sep=0.75pt]    {$D=3$};
\draw (314,62.4) node [anchor=north west][inner sep=0.75pt]    {$D=5$};
\draw (314,182.4) node [anchor=north west][inner sep=0.75pt]    {$D=3$};
\draw (414,112.4) node [anchor=north west][inner sep=0.75pt]    {$D=4$};
\draw (407,132) node [anchor=north west][inner sep=0.75pt]   [align=left] {Kerr-Sen};
\draw (228,162.4) node [anchor=north west][inner sep=0.75pt]    {$O( 2,2)$};

\end{tikzpicture}
    \caption{A schematic depiction of the series of uplifts, field redefinitions, and transformations we perform to obtain the Kerr-Sen solution.}
    \label{fig:OddProcedure}
\end{figure}

The boost procedure is elaborated on in great detail in Ref.~\cite{Hu:2025aji}, so we refer the reader there for more details. Here, we simply summarize the procedure:
\begin{enumerate}
    \item Embed the four-dimensional seed solution into five-dimensional heterotic supergravity (without gauge fields).
    \item Reduce from five dimensions to three dimensions and perform the appropriate field redefinitions to make the $O(2,2)$ symmetry manifest.
    \item Perform an $O(2,1)\subset O(2,2)$ transformation to get the boosted solution in three dimensions.
    \item Field redefine back from the $O(2,2)$-covariant frame and uplift back to five dimensions.
    \item Reduce back down to four dimensions and redefine to the $O(1,1)$ covariant frame.
    \item Truncate and field redefine to the $(\pm)$-truncation frame (without explicit derivatives of the field strengths).
\end{enumerate}
This is represented schematically in Figure \ref{fig:OddProcedure}. Note that only steps 3 and 6 particularly depend on the choice of truncation, which we will comment on below.

We will present the solution for both the $(+)$- and $(-)$-truncations. Note that the solutions are presented in the string frame. For the $(+)$-truncation, we apply the $O(2,2)$ rotation
\begin{equation}
    \Omega^{(+)}=V\,\Omega\,V^T,\label{eq:boost}
\end{equation}
where
\begin{equation}
    \Omega=\begin{pmatrix}
        \cosh\beta &\quad \sinh\beta &\quad 0 &\quad 0 \\
 \sinh\beta &\quad \cosh\beta &\quad 0 &\quad 0 \\
 0 &\quad 0 &\quad 1 &\quad 0 \\
 0  &\quad 0 &\quad 0 &\quad 1
    \end{pmatrix},\qquad V=\frac{1}{\sqrt{2}}\begin{pmatrix}
        \eta &\quad -\eta \\
 \openone &\quad \openone
    \end{pmatrix}.\label{eq:OmegaPlusTrunc}
\end{equation}
This is just the direct generalization of~\eqref{eq:KStwoDerTrans}, and it can be checked that~\eqref{eq:OmegaPlusTrunc} preserves the $(+)$-truncation. In step 6, getting from the na\"ive four-dimensional field reduction frame to the frame~\eqref{eq:wrongTrunc}, one must perform the field redefinitions
\begin{equation}
    \delta\mathcal A_{\nu}=-\frac{1}{4}e^{2\phi}\nabla^{\mu}\qty(e^{-2\phi}\mathcal F_{\mu\nu}),\qquad \delta g_{\mu\nu}=-\frac{1}{4}\mathcal F^2_{\mu\nu},\qquad \delta\phi=-\frac{1}{16}\mathcal F^2,\qquad \delta B_{\mu\nu}=\frac{1}{2}\mathcal A_{[\mu}\delta\mathcal A_{\nu]}.
\end{equation}
These arise from applying the $(+)$-truncation to~\eqref{eq:secondFieldsRedefs}. We then find the solution to be given by~\cite{Hu:2025aji}
\begin{align}
    g^{(+)}_{tt}&=-\frac{\Sigma\tilde\Delta}{\Upsilon^2}\qty(1+2\frac{\Xi}{\Upsilon}\phi)+\frac{\alpha' \mu^2}{2 \Sigma^2 \Upsilon^4}\bigg[-2 \varpi \cosh\beta \sinh ^2\frac{\beta }{2}\left(\Upsilon ^2-\mu ^2 r^2 \sinh ^2\beta \right)\nn\\
    &\quad+\left(4 a^2 r^2 x^2 (2
   \Delta -\tilde \Delta) \Sigma +2 \mu  r \Xi  \varpi +(\Delta -\mu  r) \Sigma  \varpi \right) \sinh ^2\beta \Big)\bigg],\nn\\
    g^{(+)}_{rr}&=\frac{\Sigma}{\Delta}\qty(1+2\phi)-\alpha'\frac{\mu^2 \sinh ^2\beta \left(r^2-a^2 x^2\right)^2}{2 \Delta \Sigma\Upsilon^2},\nn\\
   g^{(+)}_{xx}&=\frac{\Sigma}{1-x^2}\qty(1+2\phi)+\alpha'\frac{2 a^2 \mu^2 r^2 x^2 \sinh ^2\beta}{\left(1-x^2\right) \Sigma \Upsilon^2},\nn\\
   g^{(+)}_{ty}&=-\frac{2 a \mu r \left(1-x^2\right) \cosh ^2\frac{\beta }{2} \Sigma}{\Upsilon^2}\qty(1+2\frac{\Xi}{\Upsilon}\phi)-\frac{\alpha'\tilde\Delta\Sigma\Lambda\sinh^2\frac{\beta}{2}}{\Upsilon^2}\nn\\
   &\quad-\frac{\alpha'a\mu\sinh^2\frac{\beta}{2}}{2\Sigma^2\Upsilon^4}\bigg[\Sigma  \tilde\Delta^2 \left(2 r \Delta +\left(1-x^2\right) \left(r^2 (5 \mu -3 r)-a^2 \left(r \left(2-x^2\right)+\mu 
   x^2\right)\right)\right)\nn\\
   &\qquad\qquad+4\mu\cosh^2\frac{\beta}{2}\bigg(\Sigma ^2 \left(2 x^2 \Sigma ^2+a^2 \left(1+3 x^2-6 x^4\right) \Sigma -4 a^4 x^4 \left(1-x^2\right)\right)\nn\\
   &\qquad\qquad\quad-4 \mu  r x^2 \Sigma ^2 \left(2
   r^2-a^2 \left(-1+x^2\right)\right)+4 \mu ^2 r^2 x^2 \left(2 r^4-a^2 r^2 \left(3-5 x^2\right)+a^4 x^2 \left(1-x^2\right)\right)\bigg)\nn\\
   &\qquad\qquad + 4\mu^2r\cosh^4\frac{\beta}{2}\bigg(2 r^2 \left(r^2+a^2\right) \Delta -\left(1-x^2\right) \left(-2 a^6 x^6+r^5 (r-3 \mu )\right.\nn\\
   &\qquad\qquad\quad\left.+4 a^2 r^3 \left(r+r x^2-\mu -4 \mu  x^2\right)+a^4 r
   \left(r \left(2+2 x^2+x^4\right)+3 x^4 \mu \right)\right)\bigg)\bigg],\nn\\
   g_{yy}^{(+)}&=\frac{(1-x^2)\Sigma}{\Upsilon^2}\qty(\left(r^2+a^2\right) \Sigma +2 \mu r a^2 \left(1-x^2\right)+4 \mu r \left(r^2+a^2\right) \sinh ^2\frac{\beta }{2}+4 \mu^2 r^2 \sinh
   ^4\frac{\beta }{2})\nn\\
   &\quad +\frac{2\left(1-x^2\right) \Sigma  \phi}{\Upsilon ^3} \left[\Delta \tilde\Delta \Upsilon +4\mu r \Delta    \Upsilon  \cosh ^2\frac{\beta }{2}+4 \mu ^2 r^2 \left(\Upsilon +2 a^2
   \left(1-x^2\right)\right) \cosh ^4\frac{\beta }{2}\right.\nn\\
   &\qquad\qquad\left.-8 \mu ^2 r^2 a^2 \left(1-x^2\right) \cosh ^6\frac{\beta }{2}\right]-\frac{\alpha' a \mu r \left(1-x^2\right) \Sigma  \Lambda  \sinh ^2\beta}{\Upsilon ^2}\nn\\
   &\qquad+\frac{\alpha'a^2\mu^2(1-x^2)\sinh^2\beta}{2\Sigma^2\Upsilon^4}\bigg[\Sigma ^3 \left(2 r^2+\left(1-x^2\right) \left(a^2-\mu r\right)\right)\nn\\
   &\qquad\qquad+2 \mu  r^2 \left(2 r \left(1+x^2\right) \Sigma ^2-\left(1-x^2\right)
   \varpi  \mu \right) \sinh ^2\frac{\beta }{2}\nn\\
   &\qquad\qquad+4 \mu ^2 r^2 x^2 \left(2 r^2 \left(r^2+a^2\right)+a^2 \left(1-x^2\right)
   \left(r^2-a^2 x^2\right)\right) \sinh ^4\frac{\beta }{2}\bigg],\nn\\
   \phi^{(+)}&=-\frac{1}{2}\ln\frac{\Upsilon}{\Sigma}+\frac{\Sigma\phi}{\Upsilon}\cosh^2\frac{\beta}{2}+\alpha'\frac{\mu^2  \varpi \qty(\Xi\sinh ^2\frac{\beta }{2}-\Sigma\sinh^2\beta)}{4 \Sigma ^3 \Upsilon^2},\nn\\
   \mathcal A_t^{(+)}&=\frac{\sqrt{2}\mu r\sinh\beta}{\Upsilon}-\frac{\sqrt{2}\tilde\Delta\Sigma\sinh\beta}{\Upsilon^2}\phi\nn\\
   &\quad+\alpha'\frac{\mu^2 \sinh\beta\qty[\varpi \left(2 \mu r \sinh ^2\frac{\beta }{2}-2 \Xi +\Sigma \right)-2a^2\qty(4r^2x^2\tilde\Delta+(1-x^2)\Sigma^2)\sinh^2\frac{\beta}{2}]}{2
   \sqrt{2} \Sigma ^2 \Upsilon ^3},\nn\\
   \mathcal A_y^{(+)}&=-\frac{\sqrt{2}a \mu r (1-x^2)\sinh\beta}{\Upsilon}\qty(1+\frac{2\Sigma\cosh^2\frac{\beta}{2}}{\Upsilon}\phi)-\frac{\alpha'\Sigma\Lambda\sinh\beta}{\sqrt{2}\Upsilon}\nn\\
   &\quad+\frac{\alpha'a \mu \sinh\beta}{2\sqrt{2}\Sigma^2\Upsilon^3}\bigg[\Upsilon  \bigg(-\Sigma  \left(2 r \Delta +\left(1-x^2\right) \left(r^2 (5 \mu -3 r)-a^2 \left(r \left(2-x^2\right)+\mu 
   x^2\right)\right)\right)\nn\\
   &\qquad\quad\ \qquad\qquad\qquad+2 \mu  r^2 \left(r^2 \left(1-3 x^2\right)-a^2 x^2 \left(3-x^2\right)\right) \cosh ^2\frac{\beta
   }{2}\bigg)\nn\\
   &\qquad\qquad+2 \left(1-x^2\right) \mu  \sinh ^2\frac{\beta }{2} \left(\Delta  \Sigma ^2+2 \mu  r \left(r^2-a^2
   x^2\right)^2 \cosh ^2\frac{\beta }{2}\right)-2 \mu ^2 r \left(1-x^2\right) \varpi  \sinh ^2\beta \bigg],\nn\\
   B_{ty}^{(+)}&=\frac{2\mu a  r}{\Upsilon}(1-x^2)\sinh^2\frac{\beta}{2}\qty(1+\frac{2\Sigma}{\Upsilon}\cosh^2\frac{\beta}{2}\phi)+\frac{\alpha'\Sigma\Lambda}{\Upsilon }\cosh ^2\frac{\beta }{2}\nn\\
   &\quad+\frac{\alpha' a \mu \sinh^2\frac{\beta}{2}}{4\Sigma^2\Upsilon^3}\bigg[-2 \tilde\Delta \Sigma  \left(a^2 x^2 r \left(3-x^2\right)-r^3 \left(1-3 x^2\right)-\mu  a^2 x^2 \left(1-x^2\right)+\mu  r^2 \left(1-5
   x^2\right)\right)\nn\\
   &\qquad\qquad\qquad\qquad-8 \mu ^2 r \left(a^2 r^2 x^2 \left(1-15
   x^2\right)+2a^4 x^6+r^4(1-x^2)-\left(1+x^2\right) \varpi \right) \cosh ^4\frac{\beta }{2}\nn\\
   &\qquad\qquad\qquad\qquad+4 \mu ^2 r \left(1-x^2\right) \qty(2(r^2-a^2x^2)^2-3\varpi)  \cosh ^2\frac{\beta }{2}\bigg],\label{eq:KerrSen4derPtrunc}
\end{align}
where we have defined
\begin{equation}
   \varpi=r^4-6r^2a^2x^2+a^4x^4.
\end{equation}
We have checked that this is a solution to the $(+)$-truncation action~\eqref{eq:wrongTrunc}. This solution has a static limit given by
\begin{align}
    g^{(+)}_{tt}&=-\frac{r(r-2\mu)}{\qty(r+2\mu \sinh^2\frac{\beta}{2})^2}\qty[1-\alpha'\frac{\left(-2 \mu ^2+3 r^2+3 \mu  r\right)\mu+\left(2 \mu ^3+3 r^3+\mu ^2 r\right)\cosh\beta}{12\mu r^3\qty(r+2\mu\sinh^2\frac{\beta}{2})}],\nn\\
    g^{(+)}_{rr}&=\frac{1}{1-\frac{2\mu}{r}}\qty[1-\frac{\alpha'}{12 r}\qty(\frac{3}{\mu }+\frac{3r+4 \mu}{r^2}+\frac{6 \mu ^2 \sinh ^2\beta}{r \qty(r+2\mu \sinh^2\frac{\beta}{2})^2})],\nn\\
    g^{(+)}_{xx}&=\frac{r^2}{1-x^2}\qty[1-\frac{\alpha'}{12\mu r^3}\qty(3 r^2 + 3 r\mu + 4\mu^2)],\nn\\
    g^{(+)}_{yy}&=r^2(1-x^2)\qty[1-\frac{\alpha'}{12\mu r^3}\qty(3 r^2 + 3 r\mu + 4\mu^2)],\nn\\
    \phi^{(+)}&=-\frac{1}{2}\log\qty(1+\frac{2\mu\sinh^2\frac{\beta}{2}}{r})\nn\\
    &\quad-\frac{\alpha'}{48r^3}\qty(6 \mu +\frac{3 r^3}{\mu ^2}+\frac{3 r^2}{\mu }+10 r+\frac{6 r^2 (r-2 \mu )}{\qty(r+2\mu \sinh^2\frac{\beta}{2})^2}+\frac{r \left(14 \mu ^3-3 r^3+3 \mu  r^2-10
   \mu ^2 r\right)}{\mu ^2 \qty(r+2\mu \sinh^2\frac{\beta}{2})}),\nn\\
    \mathcal A_t^{(+)}&=\frac{\sqrt{2}\mu\sinh\beta}{r+2\mu \sinh^2\frac{\beta}{2}}+\frac{\alpha'\qty(3 r^4 - 6 r^3 \mu + r^2 \mu^2 - 10 \mu^4 + \mu (3 r^3 - 3 r^2 \mu - 14 r \mu^2 + 
    10 \mu^3)\cosh\beta)\sinh\beta}{12\sqrt{2}\mu r^2\qty(r+2\mu \sinh^2\frac{\beta}{2})^3},\label{eq:GMGHSp}
\end{align}
with $g'_{ty}$ and $B'$ vanishing. These are the four-derivative corrections to the GMGHS solution for the $(+)$-truncation.

On the other hand, the $(-)$-truncation corrections have not been previously computed. Following our discussion in Section~\ref{sec:OdpdfromOdd}, we must conjugate our boost~\eqref{eq:boost} with $U$ from~\eqref{eq:Udef}
\begin{equation}
    \Omega^{(-)}=U\,\Omega^{(+)}\,U=U\,V\,\Omega\,V^T\,U.
\end{equation}
The necessary field redefinitions from~\eqref{eq:secondFieldsRedefs} to get to the field redefinition frame~\eqref{eq:rightTrunc} in step 6 are given by
\begin{equation}
    \delta g_{\mu\nu}=-\frac{1}{4}\mathcal F^2_{\mu\nu},\qquad \delta\phi=-\frac{1}{16}\mathcal F^2.
\end{equation}
Doing so allows us to compute the $(-)$-truncated solution
\begin{align}
    g^{(-)}_{tt}&=-\frac{\Sigma\tilde\Delta}{\Upsilon^2}\qty(1+2\frac{\Xi}{\Upsilon}\phi)+\frac{\alpha' \mu^2}{2 \Sigma^2 \Upsilon^4}\bigg[-2 \varpi \cosh\beta \sinh ^2\frac{\beta }{2}\left(\Upsilon ^2-\mu ^2 r^2 \sinh ^2\beta \right)\nn\\
    &\quad+\left(4 a^2 r^2 x^2 (2
   \Delta -\tilde \Delta) \Sigma +2 \mu  r \Xi  \varpi +(\tilde\Delta+\Delta -\mu  r) \Sigma  \varpi \right) \sinh ^2\beta\bigg],\nn\\
    g^{(-)}_{rr}&=\frac{\Sigma}{\Delta}\qty(1+2\phi)-\alpha'\frac{\mu^2 \sinh ^2\beta \left(r^2-a^2 x^2\right)^2}{2 \Delta \Sigma\Upsilon^2},\nn\\
   g^{(-)}_{xx}&=\frac{\Sigma}{1-x^2}\qty(1+2\phi)+\alpha'\frac{2 a^2 \mu^2 r^2 x^2 \sinh ^2\beta}{\left(1-x^2\right) \Sigma \Upsilon^2},\nn\\
   g^{(-)}_{ty}&=-\frac{2 a \mu r \left(1-x^2\right) \cosh ^2\frac{\beta }{2} \Sigma}{\Upsilon^2}\qty(1+2\frac{\Xi}{\Upsilon}\phi)-\frac{\alpha'\tilde\Delta\Sigma\Lambda\sinh^2\frac{\beta}{2}}{\Upsilon^2}\nn\\
   &\quad-\frac{\alpha'a\mu\sinh^2\frac{\beta}{2}}{2\Sigma^2\Upsilon^4}\bigg[32 \mu ^2 r^3 a^2 x^2 \Sigma  \left(1-x^2\right) \sinh
   ^4\frac{\beta }{2}\nn\\
   &\qquad\qquad+2 \mu  \sinh ^2\frac{\beta }{2} \left(2 r^2
   \left(r^2+a^2\right) \Delta  \Upsilon +\left(1-x^2\right)
   \left(-2 \left(r^2+a^2\right) \Sigma ^3+4 r \varpi  \mu 
   \Sigma\right.\right.\nn\\
   &\qquad\qquad\qquad+\left(r^5 (-3 r+7 \mu )-2 a^2 r^3 \left(r
   \left(2+x^2\right)-2 \mu  \left(1-2 x^2\right)\right)\right.\nn\\
   &\qquad\qquad\qquad\left.\left.\left.+a^4 r
   \left(r \left(-2-2 x^2+x^4\right)+x^4 \mu \right)\right)
   \Upsilon \right)\right)\nn\\
   &\qquad\qquad-2 r \left(\Sigma ^3-3 \mu  r \Sigma ^2+\mu  r
   \left(a^2+r^2\right)^2\right) \Upsilon\nn\\
   &\qquad\qquad+\left(1-x^2\right)
   \left(4 \Sigma  \mu  \left(\left(r^2+a^2\right) \Sigma ^2-2 r
   \left(r^2-a^2 x^2\right)^2 \mu \right)\right.\nn\\
   &\qquad\qquad\qquad+\left(5 a^2 r^5
   x^2+a^4 r^3 x^4+a^2 r^4 \left(4-9 x^2\right) \mu +a^4 r^2
   \left(2+2 x^2-3 x^4\right)\right. \mu\nn\\
   &\qquad\qquad\qquad\left.\left. +24 a^2 r^3 x^2 \mu ^2-4 a^4 r
   x^4 \mu ^2+a^6 x^6 (-r+\mu )+r^5 \left(3 r^2-5 r \mu -4 \mu
   ^2\right)\right) \Upsilon \right)\bigg],\nn\\
   g_{yy}^{(-)}&=\frac{(1-x^2)\Sigma}{\Upsilon^2}\qty(\left(r^2+a^2\right) \Sigma +2 \mu r a^2 \left(1-x^2\right)+4 \mu r \left(r^2+a^2\right) \sinh ^2\frac{\beta }{2}+4 \mu^2 r^2 \sinh
   ^4\frac{\beta }{2})\nn\\
   &\quad +\frac{2\left(1-x^2\right) \Sigma  \phi}{\Upsilon ^3} \left[\Delta \tilde\Delta \Upsilon +4\mu r \Delta    \Upsilon  \cosh ^2\frac{\beta }{2}+4 \mu ^2 r^2 \left(\Upsilon +2 a^2
   \left(1-x^2\right)\right) \cosh ^4\frac{\beta }{2}\right.\nn\\
   &\qquad\qquad\left.-8 \mu ^2 r^2 a^2 \left(1-x^2\right) \cosh ^6\frac{\beta }{2}\right]+\frac{\alpha' a \mu r \left(1-x^2\right) \Sigma  \Lambda  \sinh ^2\beta}{\Upsilon ^2}\nn\\
   &\qquad+\frac{\alpha'a^2\mu^2(1-x^2)\sinh^2\beta}{2\Sigma^2\Upsilon^4}\bigg[r\Upsilon\qty(\vartheta+\mu  x^2 \left(-2 r^2 \left(a^2+r^2\right)+a^2 \left(5 r^2-a^2
   x^2\right) \left(1-x^2\right)\right))\nn\\
   &\kern13em+\Sigma  \left(\left(1-x^2\right) \left(r^2-a^2 x^2\right)^2
   \Delta +4 r^2 x^2 \left(\Upsilon +a^2
   \left(1-x^2\right)\right)^2\right)\bigg],\nn\\
   \phi^{(-)}&=-\frac{1}{2}\ln\frac{\Upsilon}{\Sigma}+\frac{\Sigma\phi}{\Upsilon}\cosh^2\frac{\beta}{2}+\alpha'\frac{\mu^2  \varpi \qty(\Xi\sinh ^2\frac{\beta }{2}-\frac{3}{2}\Sigma\sinh^2\beta)}{4 \Sigma ^3 \Upsilon^2},\nn\\
    \mathcal A_t^{(-)}&=\frac{\sqrt{2}\sinh\beta}{\Upsilon}\qty(\mu r-\frac{\tilde\Delta\Sigma}{\Upsilon}\phi+\alpha'\frac{\mu^2\varpi}{4\Sigma^2\Upsilon}),\nn\\
   \mathcal A_y^{(-)}&=-\frac{\sqrt{2}a \mu r (1-x^2)\sinh\beta}{\Upsilon}\qty(1+\frac{2\Sigma\cosh^2\frac{\beta}{2}}{\Upsilon}\phi)+\frac{\alpha'\Sigma\Lambda\sinh\beta}{\sqrt{2}\Upsilon}\nn\\
   &\quad+\frac{\alpha'a \mu \sinh\beta}{2\sqrt{2}\Sigma^2\Upsilon^2}\qty[\vartheta -\mu  \Sigma  \left(4
   r^2+\left(1-x^2\right) \left(a^2x^2 -5 r^2\right)\right)],\nn\\
   B_{ty}^{(-)}&=-\frac{2\mu a  r}{\Upsilon}(1-x^2)\sinh^2\frac{\beta}{2}\qty(1+\frac{2\Sigma}{\Upsilon}\cosh^2\frac{\beta}{2}\phi)+\frac{\alpha'\Sigma\Lambda}{\Upsilon }\cosh ^2\frac{\beta }{2}\nn\\
   &\quad+\frac{\alpha' a \mu \sinh^2\frac{\beta}{2}}{2\Sigma^2\Upsilon^2}\qty[\vartheta -\mu  \Sigma  \left(4
   r^2+\left(1-x^2\right) \left(a^2x^2 -5 r^2\right)\right)],\label{eq:KerrSen4derMtrunc}
\end{align}
where
\begin{equation}
    \vartheta=r \left(\Sigma +2 \mu  r \cosh ^2\frac{\beta
   }{2}\right) \left(2 \Sigma +\left(1-x^2\right)
   \left(a^2x^2 -3 r^2\right)\right).
\end{equation}
We have checked that this is a solution to the $(-)$-truncation action action~\eqref{eq:rightTrunc}.

Of particular note, some components have simple relations,
\begin{align}
    g^{(-)}_{rr}&=g^{(+)}_{rr},\qquad g^{(-)}_{xx}=g^{(+)}_{xx},\nn\\
    g^{(-)}_{tt}&=g^{(+)}_{tt}+\alpha'\frac{\mu^2\tilde\Delta\varpi\sinh^2\beta}{2\Sigma\Upsilon^4},\nn\\
    \phi^{(-)}&=\phi^{(+)}-\alpha'\frac{\mu^2\varpi\sinh^2\beta}{8\Sigma^2\Upsilon^2}.
\end{align}
There are no simple relations between the other components. However, it is worth noting that when comparing $g^{(+)}$, $B^{(+)}$, $\mathcal A^{(+)}$, and $\phi^{(+)}$ with $g^{(-)}$, $-B^{(-)}$, $\mathcal A^{(-)}$, and $\phi^{(-)}$, respectively, the terms proportional to $\phi$ are the same and the terms proportional to $\Lambda$ are opposite, so much of the difference between the two solutions lies in the extra terms that arise from field redefinitions. Notice also that the structure of these extra terms is the same for $\mathcal A^{(-)}_y$ and $B^{(-)}_{ty}$, which is not true for the $(+)$-truncation. Indeed, the terms are very close to one another,
\begin{equation}
    B^{(-)}_{ty}=\sqrt{2}\frac{\sinh^2\frac{\beta}{2}}{\sinh\beta}\mathcal A^{(-)}_y+\frac{\Lambda\Sigma}{\Upsilon},
\end{equation}
where the second term essentially serves to ensure the right $\beta\to 0$ limit.

The static limit for the $(-)$-truncation is given by
\begin{align}
    g^{(-)}_{tt}&=-\frac{r(r-2\mu)}{\qty(r+2\mu \sinh^2\frac{\beta}{2})^2}\left[1-\alpha'\frac{\left(-2 \mu ^2+3 r^2+3 \mu  r\right)\mu+\left(2 \mu ^3+3 r^3+\mu ^2 r\right)\cosh\beta}{12\mu r^3\qty(r+2\mu\sinh^2\frac{\beta}{2})}\right.\nn\\
    &\left.\kern11em-\alpha'\frac{\mu^2\sinh^2\beta}{2r^2\qty(r+2\mu\sinh^2\frac{\beta}{2})^2}\right],\nn\\
    g^{(-)}_{rr}&=\frac{1}{1-\frac{2\mu}{r}}\qty[1-\frac{\alpha'}{12 r}\qty(\frac{3}{\mu }+\frac{3r+4 \mu}{r^2}+\frac{6 \mu ^2 \sinh ^2\beta}{r \qty(r+2\mu \sinh^2\frac{\beta}{2})^2})],\nn\\
    g^{(-)}_{xx}&=\frac{r^2}{1-x^2}\qty[1-\frac{\alpha'}{12\mu r^3}\qty(3 r^2 + 3 r\mu + 4\mu^2)],\nn\\
    g^{(-)}_{yy}&=r^2(1-x^2)\qty[1-\frac{\alpha'}{12\mu r^3}\qty(3 r^2 + 3 r\mu + 4\mu^2)],\nn\\
    \phi^{(-)}&=-\frac{1}{2}\log\qty(1+\frac{2\mu\sinh^2\frac{\beta}{2}}{r})-\frac{\alpha'\mu^2\sinh^2\beta}{8r^2\qty(r+2\mu \sinh^2\frac{\beta}{2})^2}\nn\\
    &\quad-\frac{\alpha'}{48r^3}\qty(6 \mu +\frac{3 r^3}{\mu ^2}+\frac{3 r^2}{\mu }+10 r+\frac{6 r^2 (r-2 \mu )}{\qty(r+2\mu \sinh^2\frac{\beta}{2})^2}+\frac{r \left(14 \mu ^3-3 r^3+3 \mu  r^2-10
   \mu ^2 r\right)}{\mu ^2 \qty(r+2\mu \sinh^2\frac{\beta}{2})}),\nn\\
    \mathcal A_t^{(-)}&=\frac{\sqrt{2}\mu\sinh\beta}{r+2\mu \sinh^2\frac{\beta}{2}}-\frac{\alpha'\qty(-3 r^3+3 r^2 \mu +2 r \mu ^2+2 \mu ^3)\sinh\beta}{12\sqrt{2}\mu r^2\qty(r+2\mu \sinh^2\frac{\beta}{2})^2},\label{eq:GMGHSm}
\end{align}
and corresponds to the four-derivative corrections to the GMGHS black hole for the $(-)$-truncation.

Given the complexity of the solutions, we have explicitly checked that the solutions do indeed solve the corresponding equations of motion.

\subsection{Thermodynamics}
At the two-derivative level, the mass, charge, and angular momentum of the Kerr-Sen solution are given by
\begin{align}
    M^{(0)}&=\mu\cosh^2\frac{\beta}{2},\qquad\quad\! J^{(0)}=\mu a\cosh^2\frac{\beta}{2},\qquad Q^{(0)}=\frac{\mu}{2\sqrt{2}}\sinh\beta,\label{eq:twoDerivCharges}
\end{align}
respectively, while the horizon velocity, temperature, electric potential, and entropy are given by
\begin{align}
    \Omega^{(0)}_H&=\frac{a}{r_+^2+a^2}\sech^2\frac{\beta}{2},\nn\\
    T^{(0)}&=\frac{r_+^2-a^2}{4\pi r_+(r_+^2+a^2)}\sech^2\frac{\beta}{2},\nn\\
    \Phi_e^{(0)}&=\sqrt{2}\tanh\frac{\beta}{2},\nn\\
    S^{(0)}&=\pi(r_+^2+a^2)\cosh^2\frac{\beta}{2},
\end{align}
respectively. 

It is straightforward to compute the four-derivative corrections for the $(\pm)$-truncations. The results for the $(+)$-truncation were already computed in~\cite{Hu:2025aji}. We find the charges to be
\begin{align}
    M^{(\pm)}&=M^{(0)}+\alpha'\frac{s_{\beta}^{2}}{8\mu}\qty(1-\frac{\chi^{2}}{4}-\frac{\chi^{4}}{8}-\frac{5\chi^{6}}{64}-\frac{7\chi^{8}}{128}-\frac{21\chi^{10}}{512})+\mathcal O(\chi^{12}),\nn\\
    J^{(\pm)}&=J^{(0)}\mp\frac{5\alpha'}{32}s_{\beta}^{2}\,\chi\,\qty(1-\frac{\chi^{2}}{10}-\frac{3\chi^{4}}{80}-\frac{3\chi^{6}}{160}-\frac{7\chi^{8}}{640})+\mathcal O(\chi^{11}),\nn\\
    Q^{(\pm)}&=Q^{(0)}+\alpha'\frac{c_{\beta}s_{\beta}}{8\sqrt{2}\mu}\qty(1-\frac{\chi^{2}}{4}-\frac{\chi^{4}}{8}-\frac{5\chi^{6}}{64}-\frac{7\chi^{8}}{128}-\frac{21\chi^{10}}{512})+\mathcal O(\chi^{12}),\label{eq:fourDerivCharges}
\end{align}
where the $(\pm)$ superscript labels the charges corresponding to solutions of the $(\pm)$-truncation. In particular, notice that the mass and charge are identical for the two solutions, whereas the corrections to the angular momentum are opposite. It is straightforward to show that corrections to the mass and charge arise solely from the terms involving $\phi$, and the corrections to the angular momentum depend only on the terms not involving $\phi$. Since the $\phi$ terms are the same for the two solutions, the mass and charge are the same; on the other hand, the $\Lambda$ terms are opposite, so, naturally, the angular momentum receives opposite corrections. Note that the terms that arise from the field redefinitions are necessary; otherwise, $J$ would depend on $x$.

We also compute the corrections to the thermodynamic quantities and find that
\begin{align}
    \Omega_H^{(\pm)}&=\Omega_H^{(0)}\pm\frac{ \alpha ' as_{\beta }^2}{128 \mu ^4 c_{\beta }^4}\qty(1-\frac{5 \chi ^2}{4}-\frac{15 \chi ^4}{16}-\frac{11 \chi ^6}{16}-\frac{67 \chi ^8}{128}+\mathcal{O}(\chi^{10})),\nn\\
    T^{(\pm)}&=T^{(0)}\pm\frac{\alpha ' s_{\beta }^2}{256 \pi  \mu ^3 c_{\beta }^4}\qty(1-\frac{7 \chi ^2}{4}-\frac{7 \chi ^4}{16}-\frac{\chi ^6}{8}-\frac{3 \chi ^8}{128}+\frac{15 \chi ^{10}}{1024}+\mathcal{O}(\chi^{12})),\nn\\
    \Phi_e^{(\pm)}&=\Phi_e^{(0)}\mp\frac{\alpha ' s_{\beta }}{16 \sqrt{2} \mu ^2 c_{\beta }^3}\qty(1-\frac{3 \chi ^2}{2}-\frac{11 \chi ^4}{16}-\frac{13 \chi ^6}{32}-\frac{35 \chi ^8}{128}-\frac{51 \chi ^{10}}{256}+\mathcal{O}(\chi^{12})),\nn\\
    S^{(\pm)}&=S^{(0)}+\alpha'\frac{\pi}{2}+\frac{4\pm1}{8}\pi\alpha's_{\beta}^{2}\qty(1-\frac{5 \chi ^2}{12}-\frac{\chi ^4}{24}-\frac{\chi ^6}{192}+\frac{\chi ^8}{384}+\frac{7 \chi ^{10}}{1536}+\mathcal{O}(\chi^{12})),
\end{align}
where the entropy must be obtained by integrating the first law,\footnote{As detailed more explicitly in~\cite{Hu:2025aji}, this is due to the presence of a vertical Lorentz map in the Wald entropy that hinders a direct calculation.}
\begin{equation}
    \dd M=T\dd S+\Phi_e\dd Q+\Omega_H\dd J,\label{eq:firstLaw}
\end{equation}
and using the Wald entropy calculation for the Kerr case
\begin{equation}
    S^{(\pm)}\big\vert_{\beta=0}=S^{(0)}\big\vert_{\beta=0}+\frac{\alpha'\pi}{2}.
\end{equation}
Notice that the corrections to the horizon velocity, temperature, and electric potential are exactly opposite for the two solutions, since they do not depend on the terms involving $\phi$. On the other hand, the entropy follows a more complex pattern as it depends on all the terms in the solution.

\subsubsection{Fixing the Charges}
Note that the two-derivative values of the charges receive four-derivative corrections,~\eqref{eq:fourDerivCharges}. However, these should be fixed as the boundary conditions for the solution. Hence, we perform a parameter shift of the form
\begin{align}
    \mu&\to\mu+\frac{\alpha's_{\beta}^{2}}{8\mu c_{\beta}^{2}}\qty(1-\frac{\chi^{2}}{4}-\frac{\chi^{4}}{8}-\frac{5\chi^{6}}{64}-\frac{7\chi^{8}}{128}-\frac{21\chi^{10}}{512}+\mathcal{O}(\chi^{12})),\nn\\
    \chi&\to\chi\pm\frac{5\alpha'\chi s_{\beta}^{2}}{32\mu^{2}c_{\beta}^{2}}\qty(1-\frac{\chi^{2}}{10}-\frac{3\chi^{4}}{80}-\frac{3\chi^{6}}{160}-\frac{7\chi^{8}}{640}+\mathcal{O}(\chi^{11})),\nn\\
    \beta&\to\beta-\frac{\alpha's_{\beta}}{4\mu^{2}c_{\beta}}\qty(1-\frac{\chi^{2}}{4}-\frac{\chi^{4}}{8}-\frac{5\chi^{6}}{64}-\frac{7\chi^{8}}{128}-\frac{21\chi^{10}}{512}+\mathcal{O}(\chi^{12})),\label{eq:paramshift}
\end{align}
to maintain our boundary conditions
\bea
M^{(\pm)}=M^{(0)}+\mathcal{O}(\alpha'^{2}),\qquad J^{(\pm)}=J^{(0)}+\mathcal{O}(\alpha'^{2}),\qquad Q^{(\pm)}=Q^{(0)}+\mathcal{O}(\alpha'^{2}).
\eea
Note that the parameter shift~\eqref{eq:paramshift} depends on the choice of truncation. Then, taking~\eqref{eq:paramshift} into account, the four-derivative thermodynamic quantities for the $(+)$-truncation solution read~\cite{Hu:2025aji}
\begin{align}
    \Omega_H^{(+)}&=\Omega_{H}^{(0)}+\alpha'\frac{5as_{\beta}^{2}}{64\mu^{4}c_{\beta}^{4}}\qty(1+\frac{\chi^{2}}{5}+\frac{11\chi^{4}}{80}+\frac{\chi^{6}}{8}+\frac{77\chi^{8}}{640}+\mathcal{O}(\chi^{10})),\nn\\
    T^{(+)}&=T^{(0)}+\frac{5s_{\beta}^{2}\alpha'}{256\pi\mu^{3}c_{\beta}^{4}}\qty(1-\frac{5\chi^{2}}{4}-\frac{11\chi^{4}}{16}-\frac{\chi^{6}}{2}-\frac{263\chi^{8}}{640}-\frac{1839\chi^{10}}{5120}+\mathcal{O}(\chi^{12})),\nn\\
    \Phi_e^{(+)}&=\Phi_{e}^{(0)}-\frac{5s_{\beta}\alpha'}{16\sqrt{2}\mu^{2}c_{\beta}^{3}}\qty(1-\frac{\chi^{2}}{2}-\frac{19\chi^{4}}{80}-\frac{23\chi^{6}}{160}-\frac{63\chi^{8}}{640}-\frac{93\chi^{10}}{1280}+\mathcal{O}(\chi^{12})),\nn\\
    S^{(+)}&=S^{(0)}+\alpha'\frac{\pi}{2}+\frac{5}{8}\pi\alpha's_{\beta}^{2}\qty(1-\frac{\chi^{2}}{4}-\frac{7\chi^{4}}{40}-\frac{9\chi^{6}}{64}-\frac{77\chi^{8}}{640}-\frac{273\chi^{10}}{2560}+\mathcal{O}(\chi^{12})),
\end{align}
whereas for the $(-)$-truncation solution they read
\begin{align}
    \Omega_H^{(-)}&=\Omega_H^{(0)}-\frac{\alpha' s_{\beta }^2\,\chi }{64 \mu ^3 c_{\beta }^4}\qty(1+\chi ^2+\frac{15 \chi ^4}{16}+\frac{7 \chi ^6}{8}+\frac{105 \chi ^8}{128}+\mathcal{O}(\chi^{10})),\nn\\
    T^{(-)}&=T^{(0)}+\frac{3\alpha ' s_{\beta }^2}{256 \pi  \mu ^3 c_{\beta }^4}\qty(1+\frac{3 \chi ^2}{4}+\frac{31 \chi ^4}{48}+\frac{7 \chi ^6}{12}+\frac{69 \chi ^8}{128}+\frac{517 \chi ^{10}}{1024}+\mathcal{O}(\chi^{12})),\nn\\
    \Phi_e^{(-)}&=\Phi_e^{(0)}-\frac{3\alpha ' s_{\beta }}{16 \sqrt{2} \mu ^2 c_{\beta }^3}\qty(1+\frac{\chi ^2}{6}+\frac{\chi ^4}{16}+\frac{\chi ^6}{32}+\frac{7 \chi ^8}{384}+\frac{3 \chi ^{10}}{256}+\mathcal{O}(\chi^{12})),\nn\\
    S^{(-)}&=S^{(0)}+\alpha'\frac{\pi}{2}+\frac{3}{8}\pi\alpha'   s_{\beta }^2 \qty(1+\frac{5 \chi ^2}{12}+\frac{7 \chi ^4}{24}+\frac{15 \chi ^6}{64}+\frac{77 \chi ^8}{384}+\frac{91 \chi ^{10}}{512}+\mathcal{O}(\chi^{12})).
\end{align}
Notice that the parameter redefinition has obscured a clear relation between the corrections to the thermodynamic quantities. However, it is only after this parameter redefinition that it is reasonable to compare the two solutions; otherwise, they have different angular momenta.

\subsection{Multipole Moments}
The gravitational multipole moments are extracted using the approach of Thorne~\cite{Thorne:1980ru} by going to the asymptotically Cartesian mass-centered (ACMC-$\infty$) coordinate frame, defined by~\cite{Cano:2022wwo,Bena:2020uup}
\begin{equation}
    r_S\sqrt{1-x_S^2}=\sqrt{r^2+a^2}\sqrt{1-x^2},\qquad r_Sx_S=rx,
\end{equation}
and then expanding the metric in the far zone as
\bea
g_{tt}&=&-1+\frac{2M}{r}+\sum_{\ell\geq1}^{\infty}\frac{2}{r^{\ell+1}}\left({M}_\ell P_\ell+\sum_{\ell'<\ell}c_{\ell\ell'}^{(tt)}P_{\ell'}\right),\nn\\
g_{ty}&=&-2r(1-x^2)\left[\sum_{\ell\geq1}^{\infty}\frac{1}{r^{\ell+1}}\left(\frac{{\mathcal{S}}_\ell}{\ell}P'_\ell
+\sum_{\ell'<\ell}c_{\ell\ell'}^{(ty)}P_{\ell'}'\right)
\right],\cr
g_{rr}&=&1+\sum_{\ell\geq0}^{\infty}\frac{1}{r^{\ell+1}}\sum_{\ell'\leq\ell}c_{\ell\ell'}^{(rr)}P_{\ell'},\quad
g_{xx}=\frac{r^2}{1-x^2}\left[1+\sum_{\ell\geq0}^{\infty}\frac{1}{r^{\ell+1}}\sum_{\ell'\leq\ell}c_{\ell\ell'}^{(xx)}P_{\ell'}
\right],\nn\\
g_{yy}&=&r^2(1-x^2)\left[1+\sum_{\ell\geq0}^{\infty}\frac{1}{r^{\ell+1}}\sum_{\ell'\leq\ell}c_{\ell\ell'}^{(yy)}P_{\ell'}
\right],\quad g_{rx}=r\left[\sum_{\ell\geq0}^{\infty}\frac{1}{r^{\ell+1}}\sum_{\ell'\leq\ell}c_{\ell\ell'}^{(rx)}P_{\ell'}'
\right]\ ,
\label{AC-N}
\eea
where $P_\ell$ are the Legendre polynomials as a function of $x$ and $P_\ell'$ denotes the derivative with respect to $x$. Note that we have dropped the $S$ subscripts on the coordinates for notational convenience. $M_\ell$ and $\mathcal S_\ell$ are the mass and current multipoles, respectively, while the various $c_{\ell\ell'}$ coefficients are gauge dependent and hence unphysical~\cite{Thorne:1980ru}. Notably, $M_0$ corresponds to the mass $M$ and $\mathcal S_1$ corresponds to the angular momentum $J$. The two-derivative multipole moments are found to be
\begin{align}
    M_{2\ell}^{(0)}&=(-a^2)^\ell M^{(0)},\qquad M_{2\ell+1}^{(0)}=0,\nn\\
    \mathcal{S}_{2\ell+1}^{(0)}&=(-a^2)^\ell J^{(0)},\qquad\quad\ \mathcal{S}_{2\ell}^{(0)}=0.
\end{align}
Notice that these are independent of $\beta$ and therefore the same for both Kerr-Sen solutions and the Kerr solution.

It is then straightforward to compute the four-derivative corrected gravitational multipole moments, which we write as
\begin{equation}
    M^{(\pm)}_{\ell}=M_\ell^{(0)}+\alpha'\delta M_\ell^{(\pm)},\qquad\mathcal S^{(\pm)}_{\ell}=\mathcal S_\ell^{(0)}+\alpha'\delta \mathcal S_\ell^{(\pm)},
\end{equation}
where the $(\pm)$ superscript denotes whether we are referring to the multipole moments corresponding to the $(+)$- or $(-)$-truncation solutions.
Taking into account the shift~\eqref{eq:paramshift}, the four-derivative corrections to the gravitational multipole moments for the $(\pm)$-truncation solution are given by
\begin{align}
    \delta M_2^{(\pm)}&=-\frac{19}{120} \mu  \chi ^2   s_{\beta }^2\qty(1-\frac{55 \chi ^2}{532}-\frac{5 \chi ^4}{114}-\frac{335 \chi ^6}{13376}-\frac{525 \chi ^8}{31616})+\mathcal O(\chi^{12}),\nn\\
    \delta M_4^{(\pm)}&=\frac{333}{1960}\mu ^3 \chi ^4  s_{\beta }^2\qty(1-\frac{287 \chi ^2}{3996}-\frac{2471 \chi ^4}{87912}-\frac{46795 \chi ^6}{3047616})+\mathcal O(\chi^{12}),\nn\\
    \delta M_6^{(\pm)}&=-\frac{6863}{38808}\mu ^5 \chi ^6 s_{\beta }^2\qty(1-\frac{17087 \chi ^2}{301972}-\frac{82383 \chi ^4}{3925636})+\mathcal O(\chi^{12}),\nn\\
    \delta M_8^{(\pm)}&=\frac{2161477}{11891880}\mu ^7 \chi ^8  s_{\beta }^2\qty(1-\frac{5315009 \chi ^2}{112396804})+\mathcal O(\chi^{12}),\nn\\
    \delta \mathcal{S}_{1}^{(\pm)}&=\pm\frac{5}{32} \chi  s_{\beta }^2\qty(1-\frac{\chi^2}{10}-\frac{3 \chi^4}{80}-\frac{3 \chi^6}{160}-\frac{7 \chi^8}{640})+\mathcal O(\chi^{11}),\nn\\
    \delta \mathcal{S}_{3}^{(\pm)}&=\mp\frac{27}{160} \mu ^2 \chi^3  s_{\beta }^2\qty(1-\frac{13 \chi^2}{189}-\frac{31 \chi^4}{1296}-\frac{109 \chi^6}{9504})+\mathcal O(\chi^{11}),\nn\\
    \delta \mathcal{S}_{5}^{(\pm)}&=\pm\frac{355}{2016}\mu ^4 \chi^5  s_{\beta }^2\qty(1-\frac{23 \chi^2}{426}-\frac{223 \chi^4}{12496})+\mathcal O(\chi^{11}),\nn\\
    \delta \mathcal{S}_{7}^{(\pm)}&=\mp\frac{12439}{68640}\mu ^6 \chi^7  s_{\beta }^2\qty(1-\frac{18490 \chi^2}{410487})+\mathcal O(\chi^{11}),\nn\\
    \delta \mathcal{S}_{9}^{(\pm)}&=\pm\frac{5539}{29920}\mu ^8 \chi^9  s_{\beta }^2+\mathcal O(\chi^{11}),
\end{align}
where, as before, the result for the $(+)$-truncation was previously computed in~\cite{Hu:2025aji}. Here, we have omitted $M_0$ and $\mathcal S_1$ as these are fixed to their two-derivative values. In particular, notice that the mass multipoles are the same for the two solutions, whereas the corrections to the current multipoles are exactly opposite. Again, this is due to the fact that $\delta M_\ell$ depends only on the terms in the solution containing $\phi$, while the $\delta\mathcal S_\ell$ depend only on the terms that do not contain $\phi$. However, once we take the parameter shift~\eqref{eq:paramshift} into account, these become
\begin{align}
    \delta M_{2}^{(+)} & =-\frac{143}{240}\mu\chi^{2}s_{\beta}^{2}\qty(1-\frac{265\chi^{2}}{2002}-\frac{395\chi^{4}}{6864}-\frac{1655\chi^{6}}{50336}-\frac{5145\chi^{8}}{237952})+\mathcal O(\chi^{12}),\nn\\
    \delta M_{4}^{(+)} & =\frac{2293\mu^{3}\chi^{4}s_{\beta}^{2}}{1960}\qty(1-\frac{1981\chi^{2}}{13758}-\frac{77707\chi^{4}}{1210704}-\frac{391265\chi^{6}}{10492768})+\mathcal O(\chi^{12}),\nn \\
    \delta M_{6}^{(+)} & =-\frac{135001\mu^{5}\chi^{6}s_{\beta}^{2}}{77616}\qty(1-\frac{443975\chi^{2}}{2970022}-\frac{20776161\chi^{4}}{308882288})+\mathcal O(\chi^{12}),\nn\\
    \delta M_{8}^{(+)} & =\frac{13715861\mu^{7}\chi^{8}s_{\beta}^{2}}{5945940}\qty(1-\frac{54470591\chi^{2}}{356612386})+\mathcal O(\chi^{12}),\nn\\
    \delta\mathcal{S}_{3}^{(+)}&=-\frac{11}{20}\mu^{2}\chi^{3}s_{\beta}^{2}\qty(1-\frac{219\chi^{2}}{1232}-\frac{43\chi^{4}}{528}-\frac{743\chi^{6}}{15488})+\mathcal O(\chi^{11}),\nn \\
    \delta\mathcal{S}_{5}^{(+)} & =-\frac{557}{504}\mu^{4}\chi^{5}s_{\beta}^{2}\qty(1-\frac{1171\chi^{2}}{6684}-\frac{983\chi^{4}}{12254})+\mathcal O(\chi^{11}),\nn\\
    \delta\mathcal{S}_{7}^{(+)}&=-\frac{28529\mu^{6}\chi^{7}s_{\beta}^{2}}{17160}\qty(1-\frac{1307935\chi^{2}}{7531656})+\mathcal O(\chi^{11}),\nn\\
    \delta\mathcal{S}_{9}^{(+)}&=\frac{8307\mu^{8}\chi^{9}s_{\beta}^{2}}{3740}+\mathcal O(\chi^{11}),
\end{align}
and
\begin{align}
    \delta M_2^{(-)}&=\frac{7}{240} \mu  \chi ^2 s_{\beta }^2\qty(1+\frac{55 \chi ^2}{98}+\frac{125 \chi ^4}{336}+\frac{95 \chi ^6}{352}+\frac{345 \chi ^8}{1664})+\mathcal O(\chi^{12}),\nn\\
    \delta M_4^{(-)}&=-\frac{ 157 \mu ^3 \chi ^4 s_{\beta }^2}{1960}\qty(1+\frac{511 \chi ^2}{942}+\frac{29197 \chi ^4}{82896}+\frac{181055 \chi ^6}{718432})+\mathcal O(\chi^{12}),\nn\\
    \delta M_6^{(-)}&=\frac{10529 \mu ^5 \chi ^6 s_{\beta }^2}{77616}\qty(1+\frac{123809 \chi ^2}{231638}+\frac{8289687 \chi ^4}{24090352})+\mathcal O(\chi^{12}),\nn\\
    \delta M_8^{(-)}&=-\frac{1148989 \mu ^7 \chi ^8  s_{\beta }^2}{5945940}\qty(1+\frac{15821981 \chi ^2}{29873714})+\mathcal O(\chi^{12}),\nn\\
    \delta \mathcal{S}_3^{(-)}&=\frac{1}{20} \mu ^2 \chi ^3  s_{\beta }^2\qty(1+\frac{61 \chi ^2}{112}+\frac{17 \chi ^4}{48}+\frac{357 \chi ^6}{1408})+\mathcal O(\chi^{11}),\nn\\
    \delta \mathcal{S}_5^{(-)}&=-\frac{53}{504} \mu ^4 \chi ^5 s_{\beta }^2\qty(1+\frac{341 \chi ^2}{636}+\frac{403 \chi ^4}{1166})+\mathcal O(\chi^{11}),\nn\\
    \delta \mathcal{S}_7^{(-)}&=\frac{2789 \mu ^6 \chi ^7  s_{\beta }^2}{17160}\qty(1+\frac{390905 \chi ^2}{736296})+\mathcal O(\chi^{11}),\nn\\
    \delta \mathcal{S}_9^{(-)}&=-\frac{827 \mu ^8 \chi ^9  s_{\beta }^2}{3740}+\mathcal O(\chi^{11}).
\end{align}
Since, in both cases, the corrections to the multipole moments are nonzero, this means that we could experimentally distinguish either solution from the Kerr solution ($\beta=0$), as well as from each other.

We may also expand the gauge fields as
\begin{align}
    \mathcal A_t&=-\sum_{\ell\geq0}^{\infty}\frac{4}{r^{\ell+1}}\Big({\mathcal{Q}}_\ell P_\ell+\sum_{\ell'<\ell}c_{\ell\ell'}^{(t)}P_{\ell'}\Big)\, ,
\nn\\
\mathcal A_y&=\sum_{\ell\geq0}^{\infty}\left\{\frac{4x}{r^{2\ell}}\Big({\mathcal{P}}_{2\ell} P_{2\ell}+\sum_{\ell'<\ell}c_{2\ell,2\ell'}^{(y,1)}P_{2\ell'}\Big)+\frac{4(1-x^2)}{r^{2\ell}}\sum_{\ell'<\ell}c_{2\ell,2\ell'}^{(y,2)}P_{2\ell'}\right.\nn\\
&\qquad\qquad\left.-\frac{1-x^2}{r^{2\ell+1}}\frac{4}{2\ell+1}\Big({\mathcal{P}}_{2\ell+1} P_{2\ell+1}'+\sum_{\ell'<\ell}(c_{2\ell+1,2\ell'+1}^{(y,1)}P_{2\ell'+1}'
+c_{2\ell+1,2\ell'+1}^{(y,2)}P_{2\ell'+1})\Big)
\right\},\label{ACMC ele}
\end{align}
in order to extract the electric multipoles moments $\mathcal Q_\ell$ and the magnetic multipole moments $\mathcal P_\ell$. As before, the various $c_{\ell\ell'}$ coefficients are unphysical. Note that $\mathcal Q_0$ is always fixed to be the electric charge $Q$. At the two-derivative level, we simply find that
\begin{align}
    \mathcal Q_{2\ell}^{(0)}&=(-a^2)^\ell Q^{(0)},\qquad\ \ \mathcal Q_{2\ell+1}^{(0)}=0,\nn\\
    \mathcal{P}_{2\ell+1}^{(0)}&=-a(-a^2)^\ell Q^{(0)},\qquad \mathcal{P}_{2\ell}^{(0)}=0.
\end{align}
Note that these match the two-derivative electromagnetic multipole moments of the Kerr-Newman solution. We write the four-derivative electromagnetic multipole moments as
\begin{equation}
    Q^{(\pm)}_\ell=\mathcal Q_{\ell}^{(0)}+\alpha'\delta\mathcal Q^{(\pm)}_\ell,\qquad \mathcal P^{(\pm)}_\ell=\mathcal P_{\ell}^{(0)}+\alpha'\delta\mathcal P^{(\pm)}_\ell.
\end{equation}
We then find the four-derivative electromagnetic multipole moments for the $(\pm)$-truncation solution were found to be
\begin{align}
    \delta \mathcal{Q}_2^{(\pm)}&=-\frac{19}{120 \sqrt{2}}\mu  \chi ^2  c_{\beta } s_{\beta }\qty(1-\frac{55 \chi ^2}{532}-\frac{5 \chi ^4}{114}-\frac{335 \chi ^6}{13376}-\frac{525 \chi ^8}{31616})+\mathcal O(\chi^{12}),\nn\\
    \delta \mathcal{Q}_4^{(\pm)}&=\frac{333}{1960 \sqrt{2}}\mu ^3 \chi ^4  c_{\beta } s_{\beta }\qty(1-\frac{287 \chi ^2}{3996}-\frac{2471 \chi ^4}{87912}-\frac{46795 \chi ^6}{3047616})+\mathcal O(\chi^{12}),\nn\\
    \delta \mathcal{Q}_6^{(\pm)}&=-\frac{6863}{38808 \sqrt{2}}\mu ^5 \chi ^6  c_{\beta } s_{\beta }\qty(1-\frac{17087 \chi ^2}{301972}-\frac{82383 \chi ^4}{3925636})+\mathcal O(\chi^{12}),\nn\\
    \delta \mathcal{Q}_8^{(\pm)}&=\frac{2161477}{11891880 \sqrt{2}}\mu ^7 \chi ^8  c_{\beta } s_{\beta }\qty(1-\frac{5315009 \chi ^2}{112396804})+\mathcal O(\chi^{12}),\nn\\
    \delta \mathcal{P}_{1}^{(\pm)}&=\mp\frac{5}{32 \sqrt{2}}\chi  c_{\beta } s_{\beta }\qty(1-\frac{\chi^2}{10}-\frac{3 \chi^4}{80}-\frac{3 \chi^6}{160}-\frac{7 \chi^8}{640})+\mathcal O(\chi^{11}),\nn\\
    \delta \mathcal{P}_{3}^{(\pm)}&=\pm\frac{27}{160 \sqrt{2}}\mu ^2 \chi^3  c_{\beta } s_{\beta }\qty(1-\frac{13 \chi^2}{189}-\frac{31 \chi^4}{1296}-\frac{109 \chi^6}{9504})+\mathcal O(\chi^{11}),\nn\\
    \delta \mathcal{P}_{5}^{(\pm)}&=\mp\frac{355}{2016 \sqrt{2}}\mu ^4 \chi^5  c_{\beta } s_{\beta }\qty(1-\frac{23 \chi^2}{426}-\frac{223 \chi^4}{12496})+\mathcal O(\chi^{11}),\nn\\
    \delta \mathcal{P}_{7}^{(\pm)}&=\pm\frac{12439}{68640 \sqrt{2}}\mu ^6 \chi^7  c_{\beta } s_{\beta }\qty(1-\frac{18490 \chi^2}{410487})+\mathcal O(\chi^{11}),\nn\\
    \delta \mathcal{P}_{9}^{(\pm)}&=\mp\frac{5539}{29920 \sqrt{2}}\mu ^8 \chi^9  c_{\beta } s_{\beta }+\mathcal O(\chi^{11}),
\end{align}
where $\mathcal Q_0$ is omitted since its value is fixed to the two-derivative electric charge. As before, the $(+)$-truncation result was previously computed in~\cite{Hu:2025aji}. Once again, notice that the electric multipole moments are the same for the two truncations, while the magnetic multipole moment corrections are opposite.

Taking the parameter shift~\eqref{eq:paramshift} into account, we find
\begin{align}
    \delta\mathcal{Q}^{(+)}_{2}&=-\frac{\mu\chi^{2}s_{\beta}}{960\sqrt{2}c_{\beta}}\left(286\qty(c_{\beta}^{2}+s_{\beta}^{2})-254-\frac{5}{14}\chi^{2}\qty(106\qty(c_{\beta}^{2}+s_{\beta}^{2})-146)\right.\nn\\
    &\kern7em -\frac{5}{48}\chi^{4}\qty(158\qty(c_{\beta}^{2}+s_{\beta}^{2})-238)-\frac{5}{352}\chi^{6}\qty(662\qty(c_{\beta}^{2}+s_{\beta}^{2})-1054)\nn\\
    &\kern7em\left.-\frac{105\chi^{8}}{1664}\qty(98\qty(c_{\beta}^{2}+s_{\beta}^{2})-162)\right)+\mathcal O(\chi^{12}),\nn \\
    \delta\mathcal{Q}^{(+)}_{4} & =\frac{\mu^{3}\chi^{4}s_{\beta}}{3920\sqrt{2}c_{\beta}}\left(2293\qty(c_{\beta}^{2}+s_{\beta}^{2})-2117-\frac{7}{6}\chi^{2}\qty(283\qty(c_{\beta}^{2}+s_{\beta}^{2})-347)\right.\nonumber \\
    & \left.\kern4em-\frac{7}{528}\chi^{4}\qty(-14309+11101\qty(c_{\beta}^{2}+s_{\beta}^{2}))-\frac{245\chi^{6}}{4576}\qty(-2121+1597\qty(c_{\beta}^{2}+s_{\beta}^{2}))\right)+\mathcal O(\chi^{12}),\nn\\
    \delta\mathcal{Q}^{(+)}_{6} & =-\frac{\mu^{5}\chi^{6}s_{\beta}}{155232\sqrt{2}c_{\beta}}\left(135001\qty(c_{\beta}^{2}+s_{\beta}^{2})-126953-\frac{7}{22}\chi^{2}\qty(-73789+63425(c_{\beta}^{2}+s_{\beta}^{2}))\right.\nn\\
    &\left. \kern8em\ -\frac{21}{2288}\chi^{4}\qty(-1190837+989341\qty(c_{\beta}^{2}+s_{\beta}^{2}))\right)+\mathcal O(\chi^{12}),\nn\\
    \delta\mathcal{Q}^{(+)}_{8} & =-\frac{13040869\mu^{7}\chi^{8}s_{\beta}}{11891880\sqrt{2}c_{\beta}}\qty(1-\frac{13715861}{13040869}\qty(c_{\beta}^{2}+s_{\beta}^{2})-\frac{61475239\chi^{2}}{339062594}\qty(1-\frac{7781513}{8782177}\qty(c_{\beta}^{2}+s_{\beta}^{2})))\nn\\
    &\quad+\mathcal O(\chi^{12}),\nn\\
    \delta\mathcal{P}^{(+)}_{1} & =\frac{9\chi s_{\beta}}{32\sqrt{2}c_{\beta}}\qty(1-\frac{\chi^{2}}{6}-\frac{11\chi^{4}}{144}-\frac{13\chi^{6}}{288}-\frac{35\chi^{8}}{1152})+\mathcal O(\chi^{11}),\nn\\
    \delta\mathcal{P}^{(+)}_{3} & =\frac{\mu^{2}\chi^{3}s_{\beta}}{320\sqrt{2}c_{\beta}}\left(-182+88\qty(c_{\beta}^{2}+s_{\beta}^{2})-\frac{3}{14}\chi^{2}\qty(-137+73\qty(c_{\beta}^{2}+s_{\beta}^{2}))\right.\nn\\
    &\left. \kern6em-\frac{1}{24}\chi^{4}\qty(-323+172\qty(c_{\beta}^{2}+s_{\beta}^{2}))-\frac{1}{176}\chi^{6}\qty(-1402+743\qty(c_{\beta}^{2}+s_{\beta}^{2}))\right)+\mathcal O(\chi^{11}),\nn\\
    \delta\mathcal{P}^{(+)}_{5} & =-\frac{\mu^{4}\chi^{5}s_{\beta}}{2016\sqrt{2}c_{\beta}}\left(-1721+1114\qty(c_{\beta}^{2}+s_{\beta}^{2})-\frac{1}{6}\chi^{2}\qty(-1664+1171\qty(c_{\beta}^{2}+s_{\beta}^{2}))\right.\nn\\
    &\left.\kern7em\  -\frac{1}{176}\chi^{4}\qty(-22387+15728\qty(c_{\beta}^{2}+s_{\beta}^{2}))\right)+\mathcal O(\chi^{11}),\nn\\
    \delta\mathcal{P}^{(+)}_{7} & =\frac{\mu^{6}\chi^{7}s_{\beta}}{68640\sqrt{2}c_{\beta}}\qty(57058\qty(c_{\beta}^{2}+s_{\beta}^{2})-78077-\frac{5}{132}\chi^{2}\qty(-333007+261587\qty(c_{\beta}^{2}+s_{\beta}^{2})))+\mathcal O(\chi^{11}),\nn\\
    \delta\mathcal{P}^{(+)}_{9}&=\frac{42507\mu^{8}\chi^{9}s_{\beta}}{29920\sqrt{2}c_{\beta}}\qty(1-\frac{3692}{4723}\qty(c_{\beta}^{2}+s_{\beta}^{2}))+\mathcal O(\chi^{11}),
\end{align}
and
\begin{align}
    \delta \mathcal{Q}_2^{(-)}&=\frac{\mu  \chi ^2  s_{\beta }}{480 \sqrt{2} c_{\beta }}\Big(7 (c_{\beta }^2+s_{\beta }^2)-23+\frac{5}{14} \chi^2 \big(11 (c_{\beta }^2+s_{\beta }^2)-31\big)+\frac{25}{48} \chi ^4 \big(5 (c_{\beta }^2+s_{\beta }^2)-13\big)\nn\\
   &\qquad\qquad\qquad+\frac{35}{352} \chi ^6 \big(19 (c_{\beta}^2+s_{\beta}^2)-47\big)+\frac{105 \chi ^8 }{1664}\big(23 (c_{\beta }^2+s_{\beta }^2)-55\big)\Big)+\mathcal O(\chi^{12}),\nn\\
   \delta \mathcal{Q}_4^{(-)}&=-\frac{\mu ^3 \chi ^4  s_{\beta }}{3920 \sqrt{2} c_{\beta }}\Big(157 (c_{\beta }^2+s_{\beta }^2)-333+\frac{7}{6} \chi ^2 \big(73 (c_{\beta }^2+s_{\beta }^2)-137\big)\nn\\
   &\qquad\qquad\qquad\quad+\frac{7}{528}\chi ^4 \big(4171 (c_{\beta }^2+s_{\beta }^2)-7379\big)+\frac{245 \chi ^6 }{4576}\big(739 (c_{\beta}^2+s_{\beta}^2)-1263\big)\Big)+\mathcal O(\chi^{12}),\nn\\
   \delta \mathcal{Q}_6^{(-)}&=\frac{\mu ^5 \chi ^6  s_{\beta }}{155232 \sqrt{2} c_{\beta }}\Big(10529 (c_{\beta }^2+s_{\beta }^2)-18577+\frac{7}{22} \chi ^2 \big(17687 (c_{\beta }^2+s_{\beta}^2)-28051\big)\nn\\
   &\qquad\qquad\qquad\quad+\frac{21 \chi ^4 }{2288}\big(394747 (c_{\beta }^2+s_{\beta }^2)-596243\big)\Big)+\mathcal O(\chi^{12}),\nn\\
   \delta \mathcal{Q}_8^{(-)}&=-\frac{\mu ^7 \chi ^8  s_{\beta }}{11891880 \sqrt{2} c_{\beta }}\Big(1148989 (c_{\beta }^2+s_{\beta }^2)-1823981+\frac{7}{26} \chi ^2 \big(2260283 (c_{\beta }^2+s_{\beta
   }^2)-3260947\big)\Big)\nn\\
   &\quad+\mathcal O(\chi^{12}),\nn\\
   \delta\mathcal{P}_1^{(-)}&=-\frac{\chi   s_{\beta }}{32 \sqrt{2} c_{\beta }}\Big(1+\frac{\chi ^2}{2}+\frac{5 \chi ^4}{16}+\frac{7 \chi ^6}{32}+\frac{21 \chi ^8}{128}\Big)+\mathcal O(\chi^{11}),\nn\\
   \delta\mathcal{P}_3^{(-)}&=-\frac{\mu ^2 \chi ^3  s_{\beta }}{160 \sqrt{2} c_{\beta }}\Big(4 (c_{\beta }^2+s_{\beta }^2)-11+\frac{1}{28} \chi ^2 \big(61 (c_{\beta }^2+s_{\beta }^2)-149\big)\nn\\
   &\qquad\qquad\qquad\quad+\frac{1}{48} \chi^4 \big(68 (c_{\beta }^2+s_{\beta }^2)-157\big)+\frac{21}{352} \chi ^6 \big(17 (c_{\beta }^2+s_{\beta}^2)-38\big)\Big)+\mathcal O(\chi^{11}),\nn\\
   \delta\mathcal{P}_5^{(-)}&=\frac{\mu ^4 \chi ^5  s_{\beta }}{2016 \sqrt{2} c_{\beta }}\Big(106 (c_{\beta }^2+s_{\beta }^2)-209+\frac{1}{6} \chi ^2 \big(341 (c_{\beta }^2+s_{\beta }^2)-604\big)+\frac{1}{176}\chi ^4 \big(6448 (c_{\beta }^2+s_{\beta }^2)-10877\big)\Big)\nn\\
   &\quad+\mathcal O(\chi^{11}),\nn\\
   \delta\mathcal{P}_7^{(-)}&=-\frac{\mu ^6 \chi ^7  s_{\beta }}{68640 \sqrt{2} c_{\beta }}\Big(5578 (c_{\beta }^2+s_{\beta }^2)-9437+\frac{5}{132} \chi ^2 \big(78181 (c_{\beta }^2+s_{\beta }^2)-120017\big)\Big)+\mathcal O(\chi^{11}),\nn\\
   \delta\mathcal{P}_9^{(-)}&=-\frac{5107 \mu ^8 \chi ^9  s_{\beta }}{29920 \sqrt{2} c_{\beta }}\Big(1-\frac{3308 (c_{\beta }^2+s_{\beta }^2)}{5107}\Big)+\mathcal O(\chi^{11}).
\end{align}
Once again, the electromagnetic multipole moments are different, which means that we could, in principle, experimentally distinguish the $(+)$- and $(-)$-truncation solutions from one another. 

\subsection{Comparison with Kerr-Newman}
Comparison with the Kerr solution assumes the situation that the $U(1)$ gauge field lives in the hidden sector and would not be observable (otherwise, we could easily distinguish the two solutions based on their two-derivative electromagnetic multipole moments). However, the other possible situation is that the gauge field is the $U(1)$ of the Standard Model, in which case we should compare with the Kerr-Newman black hole solution, which is given by
\begin{align}
    \dd s_4^2&=-\frac{\mathring\Delta}{\Sigma}\left(\dd t-a(1-x^2)\dd y\right)^2+\Sigma\qty(\frac{\dd r^2}{\mathring\Delta}+\frac{\dd x^2}{1-x^2})+\frac{1-x^2}{\Sigma}\left(a\,\dd t-(r^2+a^2)\dd y\right)^2,\nn\\
    \mathcal A&=-\frac{2Qr}{\Sigma}\qty(\dd t-a(1-x^2)\dd y),
\end{align}
where
\begin{equation}
    \mathring\Delta=r^2-2M r+a^2+Q^2,\qquad \Sigma=r^2+a^2x^2.
\end{equation}
This describes a charged, rotating, stationary, axisymmetric black hole with mass $M$, electric charge $Q$, and angular momentum $J=aM$. As in~\cite{Hu:2025aji}, we wish to compare the Kerr-Sen and Kerr-Newman solutions at fixed mass, charge, and angular momentum.

We take an agnostic approach to the Einstein-Maxwell theory and consider, up to field redefinitions, the most general four-derivative action
\footnote{Following the parameterization in~\cite{Ma:2024ulp}.}
\begin{align}
    e^{-1}\mathcal L_{\mathrm{EM}}&=R-\frac{1}{4}\mathcal F^2+\alpha_1R_{\mu\nu\rho\sigma}^2+\alpha_2R_{\mu\nu\rho\sigma}\mathcal F^{\mu\nu}\mathcal F^{\rho\sigma}+\frac{\alpha_3}{4} \mathcal F^4+\frac{2\alpha_0 - 16\alpha_1 - 8\alpha_2 - 9\alpha_3}{64}(\mathcal F^2)^2\nn\\
    &\quad+\beta_1R_{\mu\nu\rho\sigma}\mathcal F^{\mu\nu}\widetilde{\mathcal F}^{\rho\sigma}+\frac{\beta_0}{4}\widetilde{\mathcal F}^{\mu\nu}\mathcal F_{\mu\nu}\mathcal F^2,
\end{align}
where $\widetilde{\mathcal F}^{\mu\nu}=\frac{1}{2}\epsilon^{\mu\nu\rho\sigma}\mathcal F_{\rho\sigma}$ is the Hodge dual. Note that the second line consists entirely of parity-odd terms. As in the Kerr case, it is not known how to compute four-derivative corrections in closed form; instead, solutions have been constructed by~\cite{Ma:2024ulp} as a perturbative expansion in
\begin{equation}
    \chi_a=\frac{a}{M},\qquad \chi_Q=\frac{Q}{M},
\end{equation} 
around $\chi_a=0$ and $\chi_Q=0$. In the case that $\chi_Q=0$, this corresponds to the perturbative expansion for the Kerr metric,~\eqref{eq:KerrExpansion}. To compare with the heterotic action, we pull out a factor of $\alpha'$,
\begin{equation}
    \alpha_i=\alpha'\bar\alpha_i,\qquad \beta_j=\alpha'\bar\beta_j.
\end{equation}
In particular, the $\bar\alpha_i$ and $\bar\beta_i$ are now just dimensionless numbers. Then the corrections to the first several multipole moments are given by~\cite{Ma:2024ulp}
\begin{align}
    \delta M_2^{(\mathrm{KN})}&=\bar\alpha_2 M \chi_a^2 \chi_Q^2-\frac{M \chi_a^2 \chi_Q^2}{300}\left[30 \bar\alpha_2 \chi_a^2+\left(8 \bar\alpha_0-76 \bar\alpha_1-203 \bar\alpha_2-19 \bar\alpha_3\right) \chi_Q^2\right]+\mathcal{O}(\chi^8),\nn\\
    \delta M_3^{(\mathrm{KN})}&=-\frac{23}{25} \bar\beta_1 M^2 \chi_a^3 \chi_Q^2+\frac{M^2 \chi_a^3 \chi_Q^2}{700}\left[91 \bar\beta_1 \chi_a^2+\qty(142 \bar\beta_0 -391 \bar\beta_1) \chi_Q^2\right]+\mathcal{O}(\chi^8),\nn\\
    \delta \mathcal{S}_2^{(\mathrm{KN})}&= \bar\beta_1 M \chi_a^2 \chi_Q^2-\frac{M \chi_a^2 \chi_Q^2}{300}\left[30 \bar\beta_1 \chi_a^2+\qty(70 \bar\beta_0-159 \bar\beta_1)\chi_Q^2\right] \nn\\
    &\quad -\frac{M \chi_a^2 \chi_Q^2}{560}\left[21 \bar\beta_1 \chi_a^4+\qty(-28\bar\beta_0+15 \bar\beta_1 )\chi_a^2 \chi_Q^2+\qty(140 \bar\beta_0 -168 \bar\beta_1)\chi_Q^4\right]+\mathcal{O}(\chi^8),\nn\\
    \delta \mathcal{S}_3^{(\mathrm{KN})}&=\frac{23}{25} \bar\alpha_2 M^2 \chi_a^3 \chi_Q^2-\frac{M^2 \chi_a^3 \chi_Q^2}{4900}\left[637\bar\alpha_2 \chi_a^2+\left(102 \bar\alpha_0-1172 \bar\alpha_1-3501 \bar\alpha_2-293 \bar\alpha_3\right) \chi_Q^2\right]\nn\\
    &\quad+\mathcal{O}(\chi^8),\nn\\
    \delta \mathcal{Q}_1^{(\mathrm{KN})}&= -\frac{1}{2} \bar\beta_1 \chi_a \chi_Q+\frac{\chi_a \chi_Q}{120}\left[6 \bar\beta_1 \chi_a^2+(3 \bar\beta_1+14 \bar\beta_0) \chi_Q^2\right]+\frac{\chi_a \chi_Q}{480}\left[4 \bar\beta_0 \chi_Q^2\left(7 \chi_Q^2-3 \chi_a^2\right)\right. \nn\\
    & \left.\quad+3 \bar\beta_1\left(3 \chi_a^4-2 \chi_a^2 \chi_Q^2+4 \chi_Q^4\right)\right]+\frac{\chi_a \chi_Q}{640}\left[\bar\beta_1\left(9 \chi_a^4 \chi_Q^2-12 \chi_a^2 \chi_Q^4+6 \chi_a^6+10 \chi_Q^6\right)\right.\nn\\
    & \left.\quad+2 \bar\beta_0\left(10 \chi_Q^6-3 \chi_a^4 \chi_Q^2\right)\right]+\mathcal{O}(\chi^8),\nn\\
    \delta \mathcal{Q}_2^{(\mathrm{KN})}&=\frac{3}{50} \bar\alpha_2 M \chi_a^2 \chi_Q+\frac{M \chi_a^2 \chi_Q}{4200}\left[\bar\alpha_2\left(90 \chi_a^2+1519 \chi_Q^2\right)-49\left(\bar\alpha_0+4 \bar\alpha_1+\bar\alpha_3\right) \chi_Q^2\right] \nn\\
    &\quad+\frac{M \chi_a^2 \chi_Q}{94080}\left[980 \bar\alpha_2 \chi_a^4-3\left(197 \bar\alpha_0+1016 \bar\alpha_1+3288 \bar\alpha_2+254 \bar\alpha_3\right) \chi_a^2 \chi_Q^2\right. \nn\\
    & \left.\qquad\qquad\quad\ \ -392\left(4 \bar\alpha_0-20 \bar\alpha_1-64 \bar\alpha_2-4 \bar\alpha_3\right) \chi_Q^4\right]+\mathcal{O}(\chi^8),\nn\\
    \delta \mathcal{P}_1^{(\mathrm{KN})}&= -\frac{1}{2} \bar\alpha_2 \chi_a \chi_Q+\frac{\chi_a \chi_Q}{120}\left[6\bar\alpha_2 \chi_a^2+\left(\bar\alpha_0-20 \bar\alpha_1-13\bar\alpha_2-5 \bar\alpha_3\right) \chi_Q^2\right]+\frac{\chi_a \chi_Q}{3360}\left[63 \bar\alpha_2 \chi_a^4\right. \nn\\
    &\quad \left.-6\left(4 \bar\alpha_0+4 \bar\alpha_1+27 \bar\alpha_2+\bar\alpha_3\right) \chi_a^2 \chi_Q^2+14\left(\bar\alpha_0-20 \bar\alpha_1-10 \bar\alpha_2-5 \bar\alpha_3\right) \chi_Q^4\right]+\mathcal{O}(\chi^8),\nn\\
    \delta\mathcal{P}_{2}^{(\mathrm{KN})}& =-\frac{3}{50}\bar\beta_{1}M\chi_{a}^{2}\chi_{Q}-\frac{M\chi_{a}^{2}\chi_{Q}}{4200}\Big[90\bar\beta_{1}\chi_{a}^{2}-49(2\bar\beta_{0}-39\bar\beta_{1})\chi_{Q}^{2}\Big] \nn\\
    &\quad-\frac{M\chi_{a}^{2}\chi_{Q}}{3360}\Big[35\bar\beta_{1}\chi_{a}^{4}-3(10\bar\beta_{0}+53\bar\beta_{1})\chi_{a}^{2}\chi_{Q}^{2}-28(13\bar\beta_{0}-30\bar\beta_{1})\chi_{Q}^{4}\Big]+\mathcal{O}(\chi^8).\label{eq:KNmultipoles}
\end{align}
Note that these multipole moments are invariant under field redefinitions.

The Kerr-Sen multipole moments we have presented above are written in terms of integration constants $\mu$, $\chi$, and $\beta$. In order to compare the solutions, we must rewrite the multipole moments in terms of physically meaningful quantities, \emph{i.e.}, the mass, charge, and angular momentum. These are related by~\cite{Hu:2025aji}
\begin{equation}
    \mu=\frac{2M}{1+\sqrt{1+8\chi_Q}},\qquad\sinh\beta=\frac{2\sqrt{2}\chi_Q}{1-2\chi_Q^2},\qquad\chi=\frac{\chi_a}{1-2\chi_Q^2}.
\end{equation}
We then rewrite the Kerr-Sen multipole moments in terms of $M$, $\chi_a$, and $\chi_Q$, and expand in $\chi_a$ and $\chi_Q$. This allows us to compare the mass multipoles
\begin{align}
    \delta M_2^{(\mathrm{KN})}&=\bar\alpha_2M\chi_a^2\chi_Q^2\qty(1-\frac{1}{10}\chi_a^2)+\mathcal O(\chi_a^5,\chi_Q^3),\nn\\
    \delta M_2^{(+)}&=-\frac{143}{120}M\chi_a^2\chi_Q^2\qty(1-\frac{265}{2002}\chi_a^2)+\mathcal O(\chi_a^5,\chi_Q^3),\nn\\
    \delta M_2^{(-)}&=\frac{7}{120}M\chi_a^2\chi_Q^2\qty(1+\frac{55}{98}\chi_a^2)+\mathcal O(\chi_a^5,\chi_Q^3),
\end{align}
the current multipoles
\begin{align}
    \delta\mathcal S_3^{(\mathrm{KN})}&=\frac{23}{25}\bar\alpha_2M^2\chi_a^3\chi_Q^2\qty(1-\frac{13}{92}\chi_a^2)+\mathcal O(\chi_a^6,\chi_Q^3),\nn\\
    \delta\mathcal S_3^{(+)}&=-\frac{11}{10}M^2\chi_a^3\chi_Q^2\qty(1-\frac{219}{1232}\chi_a^2)+\mathcal O(\chi_a^6,\chi_Q^3),\nn\\
    \delta\mathcal S_3^{(-)}&=\frac{1}{10}M^2\chi_a^3\chi_Q^2\qty(1+\frac{61}{112}\chi_a^2)+\mathcal O(\chi_a^6,\chi_Q^3),
\end{align}
the electric multipoles
\begin{align}
    \delta\mathcal Q_2^{(\mathrm{KN})}&=\frac{3}{50}\bar\alpha_2M\chi_a^2\chi_Q\qty(1+\frac{5}{14}\chi_a^2)+\mathcal O(\chi_a^5,\chi_Q^3),\nn\\
    \delta\mathcal Q_2^{(+)}&=-\frac{1}{30}M\chi_a^2\chi_Q\qty(1+\frac{25}{56}\chi_a^2)+\mathcal O(\chi_a^5,\chi_Q^3),\nn\\
    \delta\mathcal Q_2^{(-)}&=-\frac{1}{30}M\chi_a^2\chi_Q\qty(1+\frac{25}{56}\chi_a^2)+\mathcal O(\chi_a^5,\chi_Q^3),
\end{align}
and the magnetic multipoles
\begin{align}
    \delta\mathcal P_1^{(\mathrm{KN})}&=-\frac{1}{2}\bar\alpha_2\chi_a\chi_Q\qty(1-\frac{1}{10}\chi_a^2)+\mathcal O(\chi_a^3,\chi_Q^3),\nn\\
    \delta\mathcal P_1^{(+)}&=\frac{9}{32}\chi_a\chi_Q\qty(1-\frac{1}{6}\chi_a^2)+\mathcal O(\chi_a^3,\chi_Q^3),\nn\\
    \delta\mathcal P_1^{(-)}&=-\frac{1}{32}\chi_a\chi_Q\qty(1+\frac{1}{2}\chi_a^2)+\mathcal O(\chi_a^3,\chi_Q^3).
\end{align}
Thus, we see that no matter how we choose the $\bar\alpha_i$, we can never match the multipole moments of either the $(+)$- or $(-)$-truncation solutions. In particular, we can at best match one of the four multipole moments above at the leading order by a choice of $\bar\alpha_2$. Hence, the solutions are experimentally distinguishable from one another.

\section{Discussion}\label{sec:disc}
In this paper, we have found a new consistent four-derivative truncation of heterotic supergravity reduced on a torus, distinct from the vector multiplet truncation of~\cite{Liu:2023fqq}, and used it to construct higher-derivative corrected Kerr-Sen solutions. The existence of a second consistent four-derivative truncation is highly nontrivial, as the existence of a two-derivative consistent truncation does not necessarily imply the consistency of the corresponding four-derivative truncation~\cite{Liu:2023fqq}. The two truncations correspond to the two possible $O(2,1)$ invariant actions found from double field theory in~\cite{Marques:2015vua}. The solutions corresponding to either truncation have four-derivative gravitational multipole moments that are distinct from the Kerr solution and both gravitational and electromagnetic multipole moments that are distinct from the Kerr-Newman solution, regardless of the four-derivative Einstein-Maxwell action chosen. Moreover, these two Kerr-Sen solutions have distinct multipole structures from each other, so we can distinguish all these classes of black holes experimentally.

Since we can recover the gauge fields in the heterotic action by reducing on a torus and truncating, we can start in $D$ dimensions and reduce on $T^{D-d+p}$ to get a $d$-dimensional theory. We can then do the $(-)$-truncation along $T^p$ to obtain $U(1)^p$ heterotic gauge fields. This should be equivalent to the result one obtains from reducing $D$-dimensional heterotic supergravity with gauge fields on $T^{D-d}$. That is to say, we effectively extend any reduction of heterotic supergravity (without gauge fields) by promoting the generalized gauge field
\begin{equation}
    \mathbb A_\mu=\begin{pmatrix}
        A^i_\mu\\B_{\mu i}
    \end{pmatrix}\to \begin{pmatrix}
        A^i_\mu\\B_{\mu i}-\mathcal A_i^{\mathfrak a}\mathcal A^{\mathfrak a}_\mu\\
        \mathcal A_\mu^{\mathfrak a}
    \end{pmatrix},
\end{equation}
and the generalized coset as
\begin{equation}
    \mathcal H=\begin{pmatrix}
        g_{ij}-b_{ik}g^{kl}b_{lj} & \quad b_{ik}g^{kj}\\
        -g^{ik}b_{kj} & \quad g^{ij}
    \end{pmatrix}
    \to\begin{pmatrix}
        \quad g_{ij}+c_{ki}g^{kl}c_{lj}+\mathcal A_i^{\mathfrak a}\mathcal A_j^{\mathfrak a}&\quad g^{jk}c_{ki}&\quad -c_{ki}g^{kl}\mathcal A_l^{\mathfrak b}-\mathcal A_i^{\mathfrak b}\\
        g^{ik}c_{kj}&g^{ij}&\quad -g^{ik}\mathcal A_k^{\mathfrak b}\\
        -c_{kj}g^{kl}\mathcal A_l^{\mathfrak a}-\mathcal A_j^{\mathfrak a}&\quad -g^{jk}\mathcal A_k^{\mathfrak a}&\quad \delta_{\mathfrak a\mathfrak b}+\mathcal A_k^{\mathfrak a}g^{kl}\mathcal A^{\mathfrak a}_l
    \end{pmatrix},
\end{equation}
where $c_{ij}=b_{ij}-\frac{1}{2}\mathcal A_i^{\mathfrak a}\mathcal A_j^{\mathfrak a}$.

While the $(+)$-truncation always preserves half-maximal supersymmetry in any dimension, this is not necessarily the case for the $(-)$-truncation. In ten dimensions, the $(-)$-truncation is just just heterotic supergravity with $p$ $U(1)$ gauge fields, which is effectively $\mathcal N=(1,0)$ supergravity coupled to $p$ vector multiplets. Similarly, as we show in Appendix~\ref{app:susy}, the $(-)$-truncations in four dimensions is equivalent to $\mathcal N=1$ supergravity coupled to a chiral multiplet and $p$ vector multiplets. However, this cannot be true in every dimension; for example, Ref.~\cite{Cai:2025yyv} decomposed the $\mathcal N=4$ supersymmetry variations into minimal $\mathcal N=2$ ones in five dimensions, from which one sees that the $(-)$-truncation does not preserve any supersymmetry. In terms of half-maximal supermultiplets, the $F^{(-)\,a}$ form a vector multiplet with the scalars $g_{ij}$ and $b_{ij}$, the latter of which are truncated by the $(-)$-truncation. So the $(-)$-truncation can only lead to a supersymmetric theory when there are vector multiplets that do not contain scalars, which is the case for minimal supergravity in four, six, and ten dimensions, but not in five, seven, eight, or nine dimensions. Thus, since the proof should follow similarly to that for four dimensions, we conjecture that the $(-)$-truncation should preserve $\mathcal N=(1,0)$ supersymmetry in six dimensions, and we expect it to take the form of $\mathcal N=(1,0)$ supergravity coupled to a tensor multiplet and $p$ vector multiplets.

It will be interesting to consider the effect of both truncations on the extremal Kerr-Sen black hole. In this case, one can consider the near-horizon Kerr geometry, following the example of~\cite{Chen:2018jed,Cano:2023dyg,Cassani:2023vsa,Cano:2024tcr}, which found that the enhanced $SL(2,\mathbb R)\times U(1)$ symmetry in the near-horizon region reduces the equations for the scalars to ordinary differential equations, allowing one to solve exactly for the near-horizon extremal Kerr black hole. This can then be used to generate near-horizon extremal Kerr-Sen solutions. We will address this in an upcoming work.

It is worth noting that there are multiple results known for the $O(d,d)$-invariant dimensionally reduced action. In particular,~\cite{Eloy:2020dko,Elgood:2020xwu,Ortin:2020xdm,Jayaprakash:2024xlr} computed the action and associated field redefinitions from dimensional reduction, whereas~\cite{Baron:2017dvb} constructed $O(d+p,d)$-invariant actions from DFT. Our work principally used the results of~\cite{Jayaprakash:2024xlr}, which showed their action is field redefinition equivalent to the action found by~\cite{Eloy:2020dko}, but it would be interesting to check that this is also field redefinition equivalent to~\cite{Baron:2017dvb,Elgood:2020xwu,Ortin:2020xdm}. Some tentative evidence is given by the fact that the $(\pm)$-truncations of the action in~\cite{Jayaprakash:2024xlr} match the two 10D actions constructed in~\cite{Baron:2017dvb}, but it would be good to have a more complete correspondence.

\let\oldaddcontentsline\addcontentsline
\renewcommand{\addcontentsline}[3]{}
\begin{acknowledgments}
    This work is supported by the National Key Research and Development Program No. 2022YFE0134300 and the National Natural Science Foundation of China (NSFC) under Grants No. 12175164 and No. 12247103. L.M. is also supported in part by the National Natural Science Foundation of China (NSFC) grant No.~12447138, Postdoctoral Fellowship Program of CPSF Grant No.~GZC20241211, and the China Postdoctoral Science Foundation under Grant No.~2024M762338.
\end{acknowledgments}
\let\addcontentsline\oldaddcontentsline

\appendix
\section{Supersymmetry of the Truncations}\label{app:susy}
In this Appendix, we comment on the supersymmetry of the two truncated theories in four dimensions. For simplicity, we will focus on the two-derivative supersymmetry variations. Starting from heterotic supergravity in ten dimensions,~\eqref{eq:noGaugeFields}, and reducing to four dimensions on $T^6$, keeping all the fields in the reduction~\eqref{eq:redAnsatz}, we reduce the ten-dimensional gamma matrices $\Gamma^M$ as
\begin{equation}
    \Gamma^\mu=\gamma^\mu\otimes\openone,\qquad \Gamma^i=\gamma_5\otimes t^i,
\end{equation}
where the $\gamma^\mu$ form a four-dimensional Lorentzian Clifford algebra $\mathrm{Cliff}(1,3)$ and the $t^i$ form a six-dimensional Euclidean Clifford algebra $\mathrm{Cliff}(6)$. Our convention for the chiral gamma matrices is such that
\begin{equation}
    \gamma_5=i\gamma^{0123},\qquad t_*=i t^{123456}.
\end{equation}
Then we find that the ten-dimensional chiral matrix reduces as
\begin{equation}
    \Gamma^{11}=\Gamma^{0\cdots 9}=-\gamma_5\otimes t_*.\label{eq:ChiralGamma11}
\end{equation}
In particular, the supersymmetry variations in ten dimensions are with respect to a Majorana-Weyl spinor $\hat\epsilon$, such that
\begin{equation}
    \Gamma^{11}\hat\epsilon=-\hat\epsilon.\label{eq:WeylCondition10D}
\end{equation}
Inserting the decomposition~\eqref{eq:ChiralGamma11} into \eqref{eq:WeylCondition10D} requires that $\hat\epsilon$ take the form
\begin{align}
    \hat\epsilon=\epsilon_+\otimes\eta_++\epsilon_-\otimes\eta_-,\qquad \gamma_5\epsilon_\pm=\pm\epsilon_\pm,\qquad t_*\eta_\pm=\pm\eta_\pm.
\end{align}
Moreover, we expect the ten-dimensional charge conjugation matrix $\hat C$ to decompose into the four-dimensional $C_4$ and the six-dimensional $C_6$ as
\begin{equation}
    \hat C=C_4\otimes C_6.
\end{equation}
We follow the conventions that
\begin{align}
    \hat C\Gamma^M&=(\Gamma^M)^T\hat C,\qquad \hat C^*=\hat C,\qquad \hat C^T=\hat C,\qquad \hat C^2=\openone,\nn\\
    C_4\gamma^\mu&=(\gamma^\mu)^TC_4,\qquad C_4^*=C_4,\qquad C_4^T=-C_4,\qquad C_4^2=-\openone,\nn\\
    C_6t_i&=(t^i)^TC_6,\qquad C_6^*=C_6,\qquad C_6^T=-C_6,\qquad C_6^2=-\openone.
\end{align}
The charge conjugate of $\hat\epsilon$ is then given by
\begin{equation}
    \tilde{\hat\epsilon}=\tilde\epsilon_-\otimes\tilde\eta_-+\tilde\epsilon_+\otimes\tilde\eta_+,
\end{equation}
where charge conjugation of spinors is defined by
\begin{equation}
    \tilde{\hat\epsilon}:=\hat B^{-1}\hat\epsilon^*,\qquad \tilde\epsilon_{\pm}:=B_4^{-1}\epsilon_\pm^*,\qquad \tilde\eta_{\pm}:=B_6^{-1}\eta_\pm^*,
\end{equation}
and the $B$ matrices are related to the charge conjugation matrices by~\cite{VanProeyen:1999ni}
\begin{equation}
    \hat B^T=-\hat C\Gamma^0,\qquad B_4^T=-C_4\gamma^0,\qquad B_6^T=C_6,
\end{equation}
and one can easily check that $\hat B=B_4\otimes B_6$. Since $\hat\epsilon$ is Majorana-Weyl, we have a Majorana condition $\tilde{\hat\epsilon}=\hat\epsilon$. Note that charge conjugation commutes with $\Gamma^{11}$ but anticommutes with $\gamma_5$ and $t_*$, which means we identify
\begin{equation}
    \epsilon_-=\tilde\epsilon_+,\qquad \eta_-=\tilde\eta_+.
\end{equation}
Thus, our supersymmetry transformation parameter is given by
\begin{equation}
    \hat\epsilon=\epsilon_+\otimes\eta_++\tilde\epsilon_+\otimes\tilde\eta_+.
\end{equation}
The ten-dimensional supersymmetry transformations are given by
\begin{align}
    \delta_{\hat\epsilon}\psi_M&=\nabla_M(\Omega_+)\hat\epsilon,\nn\\
    \delta_{\hat\epsilon}\lambda&=\qty[\gamma^M\partial_M\phi+\frac{1}{12}H_{MNP}\gamma^{MNP}]\hat\epsilon,\label{eq:2der10DsusyVars}
\end{align}
and, upon reduction to four dimensions using the ansatz~\eqref{eq:redAnsatz}, are given by
\begin{align}
    \delta_\epsilon\psi_\mu&=\qty[\nabla_\mu(\omega_+)-\frac{1}{4}Q_\mu^{(++)\,ab}t^{ab}+\frac{1}{4}F_{\mu\nu}^{(+)\,a}\gamma^\nu\gamma_5t^a]\hat\epsilon,\nn\\
    \delta_\epsilon\tilde\lambda&=\qty[\gamma^\mu\partial_\mu\varphi+\frac{1}{12}h_{\mu\nu\rho}\gamma^{\mu\nu\rho}+\frac{1}{8}F_{\mu\nu}^{(+)\,a}\gamma^{\mu\nu}\gamma_5 t^a]\hat\epsilon,\nn\\
    \delta_\epsilon\chi^a&=\qty[-\frac{1}{2}P_\mu^{(-+)\,ab}\gamma^\mu\gamma_5t^b-\frac{1}{8}F_{\mu\nu}^{(-)\,a}\gamma^{\mu\nu}]\hat\epsilon,
\end{align}
where $\tilde\lambda=\lambda-\Gamma^i\psi_i$ is the shifted dilatino and we have defined $\chi^a=e^{ia}\psi_i$. Notice that the $t^i$ matrices act on $\eta_\pm$, which is an $\mathbf{8}$ spinor representation of $\mathfrak{so}(6)_R\simeq\mathfrak{su}(4)_R$.

Now, just from counting fields, we see that neither truncation can preserve $\mathcal N=4$ or $\mathcal N=2$ supersymmetry if we were to apply it along just one of the torus directions. However, it may be possible to preserve $\mathcal N=1$. Hence, we would like to make the $\mathcal N=1$ structure manifest. First, note that $t_*$ is Hermitian and squares to one, and so we can always find a basis for $\mathrm{Cliff}(6)$ such that
\begin{equation}
    t_*=\begin{pmatrix}
        \openone&\ 0\\
        0&\ -\openone
    \end{pmatrix}.
\end{equation}
In this basis, Dirac spinors of $\mathrm{Cliff}(6)$ can be written as
\begin{equation}
    \eta=\begin{pmatrix}
        \bar\eta_+^{\mathcal A}\\ \bar\eta_{-{\mathcal A}}
    \end{pmatrix},
\end{equation}
where the $\mathcal A$ index runs from 1 to 4 and the bar denotes four-component Weyl spinors. Then we have that the $t^i$ are Hermitian and anticommute with $t_*$, which forces them to have the form
\begin{equation}
    t^i=\begin{pmatrix}
        0&\ (X^{\dagger i})^\mathcal{AB}\\ (X^i)_\mathcal{AB}&\ 0
    \end{pmatrix}.
\end{equation}
The charge conjugation matrix can be written in the form
\begin{equation}
    C_6=\begin{pmatrix}
        0&\ \mathcal C\\\mathcal C&\ 0
    \end{pmatrix},\qquad\mathcal C^*=\mathcal C,\qquad\mathcal C^T=-\mathcal C,\qquad \mathcal C^2=-\openone.
\end{equation}
One can check that this satisfies $\mathcal C X^i=(X^i)^T\mathcal C$, which means that $\mathcal C$ acts as a charge conjugation matrix on the four-dimensional subspace. The charge conjugate of $X^i$ is then given by
\begin{equation}
    \tilde X^i:=\mathcal C(X^i)^*\mathcal C^{-1}=(X^i)^\dagger,
\end{equation}
and so we may rewrite the $t^i$ as
\begin{equation}
    t^i=\begin{pmatrix}
        0&\ (\tilde X^{i})^\mathcal{AB}\\ (X^i)_\mathcal{AB}&\ 0
    \end{pmatrix}.
\end{equation}
Similarly, we may use the $t^i$ to construct $t^{ij}$ as
\begin{equation}
    t^{ij}=\begin{pmatrix}
        (X^{ij})^\mathcal{A}{}_{\mathcal B}&\ 0\\0&\ (\tilde X^{ij})_\mathcal{A}{}^{\mathcal B}
    \end{pmatrix},\qquad X^{ij}=\tilde X^{[i} X^{j]}.
\end{equation}
Here, $\tilde X^{ij}$ is the charge conjugate of $X^{ij}$, which itself is built out of $X^i$ and its charge conjugate $\tilde X^i$. Note that the $t^{ij}$ are generators of $\mathfrak{so}(6)_R\simeq \mathfrak{su}(4)_R$, so we see that this parametrization decomposes the reducible $\mathbf{8}$ representation into a $\mathbf{4}$ irrep and its conjugate $\bar{\mathbf{4}}$. In particular, $\bar\eta_+^\mathcal{A}$ transforms in the $\mathbf{4}$ and $\tilde{\bar\eta}_{+\mathcal{A}}$ transforms in the $\bar{\mathbf{4}}$. We see that $\mathcal C$ is used to raise and lower indices using the NW-SE convention
\begin{equation}
    \bar\eta_\mathcal{A}=\bar\eta^{\mathcal B}\mathcal C_\mathcal{BA},\qquad\bar\eta^\mathcal{A}=\mathcal C^\mathcal{AB}\bar\eta_{\mathcal B},
\end{equation}
as we would expect~\cite{Freedman:2012zz}.

We would like to select a $\mathfrak{su}(3)$ invariant subsector to make the $\mathcal N=1$ supersymmetry manifest. Without loss of generality, we will choose $\eta_+^2=\eta_+^3=\eta_+^4=0$. Then the gravitino variation decomposes as
\begin{align}
    \delta_\epsilon\psi_\mu^\mathcal{A}&=\qty[\nabla_\mu(\omega_+)\delta^{\mathcal A}_{\mathcal B}-\frac{1}{4}Q_\mu^{(++)\,ab}(X^{ab})^{\mathcal A}{}_{\mathcal B}]\eta_+^{\mathcal B}\epsilon_+-\frac{1}{4}F_{\mu\nu}^{(+)\,a}(\tilde X^a)^{\mathcal{AB}}\tilde\eta_{+\mathcal B}\gamma^\nu\tilde\epsilon_+,\nn\\
    \delta_\epsilon\psi_{\mu\mathcal A}&=\qty[\nabla_\mu(\omega_+)\delta_{\mathcal A}^{\mathcal B}-\frac{1}{4}Q_\mu^{(++)\,ab}(\tilde X^{ab})_{\mathcal A}{}^{\mathcal B}]\tilde\eta_{+\mathcal B}\tilde\epsilon_++\frac{1}{4}F_{\mu\nu}^{(+)\,a}(X^a)_{\mathcal{AB}}\eta_{+}^{\mathcal B}\gamma^\nu\epsilon_+.
\end{align}
Note that these two equations are just charge conjugates of one another (with respect to $\hat C=C_4\otimes C_6$). From the four-dimensional perspective, $\epsilon_+$ is left-handed while $\tilde\epsilon_+$ is right-handed. We expect the $\mathcal N=1$ supersymmetry algebra to consist of left-handed gravitini transforming with a left-handed $\epsilon$~\cite{Cremmer:1982en}.

Now define
\begin{equation}
    \xi_{\mathcal A}=\begin{pmatrix}
        1&\ 0&\ 0&\ 0
    \end{pmatrix},\qquad \xi^{\mathcal A}=\mathcal C^{\mathcal{AB}}\xi_{\mathcal B}=\mathcal C^{\mathcal A1}.
\end{equation}
In particular, since $\mathcal CX^i$ is antisymmetric,\footnote{This can be seen by using the fact that $\mathcal C X^i=(X^i)^T\mathcal C$.} this will satisfy
\begin{equation}
    \xi^{\mathcal A}(X^a)_{\mathcal{A}1}=0.
\end{equation}
Then we may define a projector by
\begin{equation}
    \mathfrak P^{\mathcal A}{}_{\mathcal B}=\delta^{\mathcal A}_{\mathcal B}-\xi^{\mathcal A}\xi_{\mathcal B},
\end{equation}
and denote indices projected with $\mathfrak P$ by a bar. For example,
\begin{equation}
    \psi_\mu^{\bar{\mathcal A}}:=\mathfrak{P}^{\mathcal A}{}_{\mathcal B}\psi_\mu^{\mathcal B}.
\end{equation}
Then the gravitino variation can be decomposed into
\begin{align}
    \delta_\varepsilon\qty(\xi_{\mathcal A}\psi_\mu^\mathcal{A})&=\xi_{\mathcal A}\qty[\nabla_\mu(\omega_+)\delta^{\mathcal A}_{\mathcal B}-\frac{1}{4}Q_\mu^{(++)\,ab}(X^{ab})^{\mathcal A}{}_{\mathcal B}]\varepsilon_+^{\mathcal B},\nn\\
    \delta_{\varepsilon}\psi_{\mu\bar{\mathcal{A}}}&=\frac{1}{4}F_{\mu\nu}^{(+)\,a}( X^a)_{\mathcal{AB}}\gamma^\nu\varepsilon_+^{\mathcal B},
\end{align}
where we have defined the four-dimensional spinor
\begin{equation}
    \varepsilon_+^{\mathcal A}:=\epsilon_+\eta_+^{\mathcal A},
\end{equation}
and we now view $\mathcal A$ as an $R$-symmetry index associated with that spinor, and our restriction of $\eta_+$ to have one component is then viewed as a restriction of the $R$-symmetry to a $U(1)_R$ sector, appropriate for $\mathcal N=1$. We then interpret this as the gravity multiplet $(g_{\mu\nu},\xi_\mathcal{A}\psi_\mu^{\mathcal A})$ and three gravitino multiplets $(\psi_{\mu\bar{\mathcal{A}}},F_{\mu\nu}^{(+)\,a}(X^a)_{\mathcal A1})$.

A similar story holds for the other supersymmetry variations, with the dilatino giving rise to
\begin{align}
    \delta(\xi_{\mathcal A}\tilde\lambda^{\mathcal A})&=\qty[\gamma^\mu\partial_\mu\varphi+\frac{1}{12}h_{\mu\nu\rho}\gamma^{\mu\nu\rho}]\xi_{\mathcal A}\varepsilon_{+}^{\mathcal A},\nn\\
    \delta\tilde\lambda^{\bar{\mathcal{A}}}&=-\frac{1}{8}F_{\mu\nu}^{(+)\,a}(\tilde X^a)^{\mathcal{AB}}\gamma^{\mu\nu}\tilde\varepsilon_{+\mathcal B}.
\end{align}
Note that the three-form flux $h$ dualizes to an axion in four dimensions. Hence, we interpret this as a chiral multiplet $(\xi_{\mathcal A}\tilde\lambda^{\mathcal A},\varphi+i\star h)$ and three vector multiplets $(F^{(+)\,a}_{\mu\nu}(\tilde X^a)^{\mathcal{AB}}\mathcal C_{\mathcal B1},\tilde\lambda^{\bar A})$. Finally, the gaugini become
\begin{align}
    \delta_{\varepsilon}\qty(\xi^{\mathcal A}\chi^a_{\mathcal A})&=-\frac{1}{8}F^{(-)\,a}_{\mu\nu}\gamma^{\mu\nu}\xi^{\mathcal A}\tilde\varepsilon_{+\mathcal A},\nn\\
    \delta_\varepsilon\chi^a_{\bar A}&=-\frac{1}{2}P_\mu^{(-+)\,ab}(X^b)_{\mathcal A\mathcal B}\gamma^\mu\varepsilon^{\mathcal B}_+.
\end{align}
We then interpret this as six vector multiplets $(F_{\mu\nu}^{(-)\,a},\xi^{\mathcal A}\chi^a_{\mathcal A})$ and 18 chiral multiplets $(\chi^a_{\bar{\mathcal A}},P_\mu^{(-+)\,ab}(X^b)_{\mathcal{A}1})$.

Thus, in the $\mathcal N=1$ language, the theory can be written as the $\mathcal N=1$ supergravity multiplet $(g_{\mu\nu},\xi_\mathcal{A}\psi_\mu^{\mathcal A})$ coupled to three gravitino multiplets $(\psi_{\mu\bar{\mathcal{A}}},F_{\mu\nu}^{(+)\,a}(X^a)_{\mathcal A1})$, nine vector multiplets $(F^{(+)\,a}_{\mu\nu}(\tilde X^a)^{\mathcal{AB}}\mathcal C_{\mathcal B1},\tilde\lambda^{\bar{\mathcal{A}}})$ and $(F_{\mu\nu}^{(-)\,a},\xi^{\mathcal A}\chi^{a}_{\mathcal A})$, and 19 chiral multiplets $(\xi_{\mathcal A}\tilde\lambda^{\mathcal A},\varphi+i\star h)$ and $(\chi^a_{\bar{\mathcal A}},P_\mu^{(-+)\,ab}(X^b)_{\mathcal{A}1})$. Note that this is just one possible realization of the $\mathcal N=1$ subalgebra of the $\mathcal N=4$ algebra. However, we see that applying the $(-)$-truncation along $n$ of the 6 compactified directions and the $(0)$-truncation along the rest (\emph{i.e.}, truncating $F^{(+)\,a}$, $g_{ij}$, and $b_{ij}$, but keeping $n$ of the $F^{(-)\,a}$) corresponds to removing the three gravitino multiplets, $9-n$ of the nine vector multiplets, and 18 of the 19 chiral multiplets, leaving behind $\mathcal N=1$ supergravity coupled to $n$ vector multiplets and one chiral multiplet. In particular, the version of the $(-)$-truncation to which the Kerr-Sen metric is a solution, where we keep just one of the $F^{(-)\,i}$, corresponds to $\mathcal N=1$ supergravity coupled to one vector multiplet and one chiral multiplet. Note also that the $(0)$-truncation, to which the Kerr metric is a solution, corresponds to $\mathcal N=1$ supergravity coupled to a single chiral multiplet. Nevertheless, the solutions themselves are not, in general, supersymmetric, with the possible exception of the extremal limit of GMGHS~\cite{Kallosh:1992ii}.

On the other hand, it is not clear that the $(+)$-truncation preserves even $\mathcal N=1$ supersymmetry. Of course, keeping all the $F^{(+)\,a}$ and truncating the $F^{(-)\,a}$, $g_{ij}$, and $b_{ij}$ corresponds to $\mathcal N=4$ supergravity coupled to 6 vector multiplets, but issues arise when we wish to keep only one of the $F^{(+)\,a}$, as we do when working with the Kerr-Sen solution. The difficulty comes down to disentangling the gravitino multiplets $(\psi_{\mu\bar{\mathcal{A}}},F_{\mu\nu}^{(+)\,a}(X^a)_{\mathcal A1})$ from the vector multiplets $(F^{(+)\,a}_{\mu\nu}(\tilde X^a)^{\mathcal{AB}}\mathcal C_{\mathcal B1},\tilde\lambda^{\bar{\mathcal{A}}})$. This is most easily addressed by explicitly choosing a basis for our Clifford algebra. Let $\bar t^i$ denote the Weyl-Brauer matrices
\begin{align}
    \bar t^1&=\sigma^1\otimes\openone\otimes\openone,\qquad\bar t^2=\sigma^2\otimes\openone\otimes\openone,\qquad\bar t^3=\sigma^3\otimes\sigma^1\otimes\openone,\nn\\
    \bar t^4&=\sigma^3\otimes\sigma^2\otimes\openone,\qquad\bar t^5=\sigma^3\otimes\sigma^3\otimes\sigma^1,\qquad\bar t^6=\sigma^3\otimes\sigma^3\otimes\sigma^2,
\end{align}
where the $\sigma^i$ are the Pauli matrices. Then we will take $t^i$ to be given by
\begin{equation}
    t^i=\mathfrak B^T\bar t^i\mathfrak B,\qquad\mathfrak B=\begin{pmatrix}
         1 & \quad 0 & \quad 0 & \quad 0 & \quad 0 & \quad 0 & \quad 0 & \quad 0 \\
         0 & \quad 0 & \quad 0 & \quad 0 & \quad 0 & \quad 0 & \quad 1 & \quad 0 \\
         0 & \quad 0 & \quad 0 & \quad 0 & \quad 0 & \quad 1 & \quad 0 & \quad 0 \\
         0 & \quad 0 & \quad 0 & \quad 1 & \quad 0 & \quad 0 & \quad 0 & \quad 0 \\
         0 & \quad 0 & \quad 0 & \quad 0 & \quad 1 & \quad 0 & \quad 0 & \quad 0 \\
         0 & \quad 0 & \quad 1 & \quad 0 & \quad 0 & \quad 0 & \quad 0 & \quad 0 \\
         0 & \quad 1 & \quad 0 & \quad 0 & \quad 0 & \quad 0 & \quad 0 & \quad 0 \\
         0 & \quad 0 & \quad 0 & \quad 0 & \quad 0 & \quad 0 & \quad 0 & \quad 1
    \end{pmatrix}.
\end{equation}
We choose the charge conjugation matrix to be given by
\begin{equation}
    C_6=i\mathfrak B^T\qty(\sigma^1\otimes\sigma^2\otimes\sigma^1)\mathfrak B.
\end{equation}
In this basis, the two combinations of field strengths become
\begin{equation}
    F^{(+)\,a}(X^a)_{\mathcal A1}=\begin{pmatrix}
        F^{(+)\,1}-iF^{(+)\,2}\\-F^{(+)\,3}-iF^{(+)\,4}\\-F^{(+)\,5}-iF^{(+)\,6}\\0
    \end{pmatrix},\qquad F^{(+)\,a}(\tilde X^a)^{\mathcal{AB}}\mathcal C_{\mathcal B1}=\begin{pmatrix}
        0\\F^{(+)\,5}-iF^{(+)\,6}\\-F^{(+)\,3}+iF^{(+)\,4}\\-F^{(+)\,1}-iF^{(+)\,2}
    \end{pmatrix}.
\end{equation}
Thus, if we keep one $F^{(+)\,a}$, say $F^{(+)\,1}$, and truncate the rest, then it is ``shared'' between a gravitino multiplet and a vector multiplet, rather than allowing us to set one of the multiplets to zero. Note that there is no change of basis that one can perform to remove this issue. Hence, it is not clear that the resulting theory preserves any supersymmetry.

\subsection{Four-derivative Consistency of Supersymmetry}
Now that we know the multiplet structure of the two-derivative theory, checking that the four-derivative truncation is consistent with supersymmetry amounts to checking that the extra four-derivative terms that appear do not disrupt the structure of the supersymmetry variations. The $(+)$-truncation, where we keep all six vector fields $F^{(+)\,a}$, has been shown to preserve $\mathcal N=4$ supersymmetry at the four-derivative level~\cite{Liu:2023fqq}, whereas, as we have just seen, keeping just one of the $F^{(+)\,a}$ does not preserve any supersymmetry, even at the two-derivative level. Hence, we will focus on the $(-)$-truncation. The four-derivative supersymmetry variations are given by~\cite{Bergshoeff:1989de}
\begin{align}
    \delta_{\hat\epsilon}\psi_M&=\nabla_M(\Omega_+)\hat\epsilon,\nn\\
    \delta_{\hat\epsilon}\lambda&=\qty[\gamma^M\partial_M\phi+\frac{1}{12}\tilde H_{MNP}\gamma^{MNP}]\hat\epsilon,\label{eq:4der10DsusyVars}
\end{align}
which is just the replacement $H\to\tilde H$ in~\eqref{eq:2der10DsusyVars}. Upon reduction on $T^6$ using the ansatz~\eqref{eq:redAnsatz}, the supersymmetry variations~\eqref{eq:4der10DsusyVars} become~\cite{Jayaprakash:2024xlr}
\begin{align}
    \delta_\epsilon\psi_\mu&=\Bigl[\nabla_\mu(\tilde\omega_+)-\fft14Q_\mu^{(++)\,ab}t^{ab}+\fft{1}4F_{\mu\nu}^{(+)\,a}\qty(\delta^{ab}-\alpha'T^{ab})\gamma^\nu\gamma_5 t^b\nn\\
    &\quad+\fft{\alpha'}8\Bigl(\Bigl(\fft12R_{\mu\nu\alpha\beta}(\tilde\omega_-)F_{\alpha\beta}^{(+)\,a}+2P_\alpha^{(-+)\,ba}\mathcal D_{[\mu}^{\prime(-)}F_{\nu]\alpha}^{(-)\,b}-\fft14F_{\alpha\beta}^{(+)\,a}F_{\beta[\mu}^{(-)\,b}F_{\nu]\alpha}^{(-)\,b}\Bigr)\gamma^\nu\gamma_5 t^a\nn\\
    &\kern3em+\Bigl(P_\alpha^{(-+)\,ca}\mathcal D_\mu^{(-)}P_\alpha^{(-+)\,cb}+\fft18F_{\alpha\beta}^{(+)\,a}\mathcal D_\mu^{(-)}F_{\alpha\beta}^{(+)\,b}-\fft12F_{\alpha\beta}^{(+)\,a}F_{\mu\alpha}^{(-)\,c}P_\beta^{(-+)\,cb}\Bigr)t^{ab}\Bigr)\Bigr]\hat\epsilon,\nn\\
    \delta_\epsilon\tilde\lambda&=\Bigl[\gamma^\mu\partial_\mu\varphi+\fft{1}{12}\tilde h_{\mu\nu\rho}\gamma^{\mu\nu\rho}+\fft18F_{\mu\nu}^{(+)\,a}\qty(\delta^{ab}-\alpha'T^{ab})\gamma^{\mu\nu}\gamma_5t^b\nn\\
    &\quad+\fft{\alpha'}{16}\Bigl(\Bigl(\fft12R_{\mu\nu\alpha\beta}(\tilde\omega_-)F_{\alpha\beta}^{(+)\,a}+2P_\alpha^{(-+)\,ba}\mathcal D_{[\mu}^{\prime(-)}F_{\nu]\alpha}^{(-)\,b}-\fft14F_{\alpha\beta}^{(+)\,a}F_{\beta\mu}^{(-)\,b}F_{\nu\alpha}^{(-)\,b}\Bigr)\gamma^{\mu\nu}\gamma_5t^a\nn\\
    &\kern3em+\Bigl(\fft1{12}F_{\alpha\beta}^{(+)\,a}F_{\beta\gamma}^{(+)\,b}F_{\gamma\alpha}^{(+)\,c}-F_{\alpha\beta}^{(+)\,a}P_\beta^{(-+)\,db}P_\alpha^{(-+)\,dc}\Bigr)\gamma_5t^{abc}\Bigr)\Bigr]\hat\epsilon,\nn\\
    \delta_\epsilon\chi^a&=\left[-\fft{1}2\gamma^\mu P_\mu^{(-+)\,ab}\qty(\delta^{bc}-\alpha'T^{bc})\gamma_5t^c-\fft18F_{\mu\nu}^{(-)\,a}\gamma^{\mu\nu}\right]\hat\epsilon,
\label{eq:redsusy}
\end{align}
where
\begin{equation}
    \tilde{\omega}_{\pm\mu}^{\alpha\beta}=\omega_\mu^{\alpha\beta}\pm\frac{1}{2}\tilde h_\mu{}^{\alpha\beta},\qquad T^{ab}=\fft1{32}F_{\mu\nu}^{(+)\,a}F^{(+)\,\mu\nu\,b}+\fft14P_\mu^{(-+)\,ca}P^{(-+)\,\mu\,cb}.
\end{equation}
Here, $\chi^a=e^{ia}\tilde\psi_i$, where the shifted gaugino is given by
\begin{equation}
    \tilde{\psi}_i=\psi_i+\alpha'e_i^a\left[-\frac{1}{4} F_{\mu \nu}^{(+)\, a} \hat{\mathcal{D}}_\mu \psi_\nu-\frac{1}{2} P_\mu^{(-+)\, b a} e_b^j\left(\hat{\mathcal{D}}_\mu \psi_j-\hat{\mathcal{D}}_j \psi_\mu\right)-\frac{1}{16} F_{\mu \nu}^{(+)\, a}\left(F_{\mu \nu}^{(+)\, b}+F_{\mu \nu}^{(-)\, b}\right) e_b^j \psi_j\right],
\end{equation}
where
\begin{align}
    \hat{\mathcal{D}}_\mu & =\nabla_\mu\qty(\omega_{+})-\frac{1}{4} Q^{(++)\, a b} \gamma^{a b}+\frac{1}{4} F_{\mu \nu}^{(+)\, a} \gamma^\nu \gamma_5 t^a, \nn\\
    \hat{\mathcal{D}}_i &=e_i^a\left(-\frac{1}{2} P_\mu^{(-+)\, a b} \gamma^\mu \gamma_5 t^b-\frac{1}{8} F_{\mu \nu}^{(-)\, a} \gamma^{\mu \nu}\right).
\end{align}
Note that the supersymmetry variations~\eqref{eq:redsusy} are in the field redefinition frame that precedes the application of~\eqref{eq:secondFieldsRedefs}. Applying the $(-)$-truncation~\eqref{eq:secondTrunc} to the supersymmetry variations~\eqref{eq:redsusy}, these simply become
\begin{align}
    \delta_\epsilon\psi_\mu&=\nabla_\mu(\omega_+)\hat\epsilon,\nn\\
    \delta_\epsilon\tilde\lambda&=\qty[\gamma^\mu\partial_\mu\varphi+\frac{1}{12}\tilde h_{\mu\nu\rho}\gamma^{\mu\nu\rho}]\hat\epsilon,\nn\\
    \delta_\epsilon\chi^a&=-\frac{1}{4\sqrt{2}}\mathcal F_{\mu\nu}^{a}\gamma^{\mu\nu}\hat\epsilon,
\end{align}
Importantly, the four-derivative supersymmetry variations have the same structure as the two-derivative ones, with the only change being the promotion of $h$ to $\tilde h$. Note that $\tilde h$ dualizes to the axion in four dimensions. Thus, the supermultiplet structure will remain unchanged, and we see that the four-derivative $(-)$-truncation is consistent with $\mathcal N=1$ supersymmetry. Moreover, note that field redefinitions should not affect the consistency of the supersymmetry algebra, and hence the supermultiplet structure should remain after the field redefinitions~\eqref{eq:secondFieldsRedefs} are performed.

\let\oldaddcontentsline\addcontentsline
\renewcommand{\addcontentsline}[3]{}
\bibliography{cite}
\let\addcontentsline\oldaddcontentsline

\end{document}